\documentclass[journal,twocolumn]{IEEEtran}

\usepackage{cite}
\usepackage{url}

\usepackage[dvipsnames,svgnames]{xcolor}
\usepackage{graphicx}
\usepackage{psfrag}
\DeclareGraphicsExtensions{.xps}
\usepackage{epstopdf} 
\usepackage{subcaption}
\usepackage[font=small,labelfont=bf]{caption}
\usepackage[cmex10]{amsmath}
\usepackage{amssymb}
\usepackage{amsthm}

\def\tr{\mathrm{tr}}
\def\vecoperator{\mathrm{vec}}

\def\diag{\mathrm{diag}}

\newtheorem{theorem}{Theorem}
\newtheorem{lemma}{Lemma}
\newtheorem{corollary}{Corollary}
\newtheorem{example}{Example}
\newtheorem{remark}{Remark}
\newtheorem{assumption}{Assumption}

\usepackage[textwidth=30mm]{todonotes}

\newcommand{\vect}[1]{\mathbf{#1}}

\def\diag{\mathrm{diag}}

\def\tr{\mathrm{tr}}

\def\Htran{\mbox{\tiny $\mathrm{H}$}}
\def\Ttran{\mbox{\tiny $\mathrm{T}$}}
\def\CN{\mathcal{N}_{\mathbb{C}}} 
\def\taupu{\tau_{p}} 
\def\bphiu{\boldsymbol{\phi}} 

\def\lalt{l'} 

\begin{document}

\IEEEoverridecommandlockouts

\title{Massive MIMO Has Unlimited Capacity}

\author{
\IEEEauthorblockN{Emil Bj{\"o}rnson, \emph{Member, IEEE},
Jakob Hoydis, \emph{Member, IEEE}, Luca Sanguinetti, \emph{Senior Member, IEEE}
\thanks{
\copyright 2017 IEEE. Personal use of this material is permitted. Permission from IEEE must be obtained for all other uses, in any current or future media, including reprinting/republishing this material for advertising or promotional purposes, creating new collective works, for resale or redistribution to servers or lists, or reuse of any copyrighted component of this work in other works.
\newline \indent E.~Bj\"ornson is with the Department of Electrical Engineering (ISY), Link\"{o}ping University, 58183 Link\"{o}ping, Sweden (emil.bjornson@liu.se).  J.~Hoydis is with Nokia Bell Labs, Paris-Saclay, 91620 Nozay, France (jakob.hoydis@nokia-bell-labs.com). L.~Sanguinetti is with the University of Pisa, Dipartimento di Ingegneria dell'Informazione, 56122 Pisa Italy (luca.sanguinetti@unipi.it) and also with the Large Systems and Networks Group (LANEAS), CentraleSup\'elec, Universit\'e Paris-Saclay, 3 rue Joliot-Curie,  91192 Gif-sur-Yvette, France. The authors have contributed equally to this work and are listed alphabetically. 
\newline\indent This research has been supported by ELLIIT, the Swedish Foundation for Strategic Research (SFF), the EU FP7 under ICT-619086 (MAMMOET), and the ERC Starting Grant 305123 MORE.
\newline\indent Parts of this paper were presented at the International Conference on Communications (ICC), 21--25 May, 2017, Paris, France.}
}}
\maketitle

\begin{abstract}
The capacity of cellular networks can be improved by the unprecedented array gain and spatial multiplexing offered by Massive MIMO. Since its inception, the coherent interference caused by pilot contamination has been believed to create a finite capacity limit, as the number of antennas goes to infinity. In this paper, we prove that this is incorrect and an artifact from using simplistic channel models and suboptimal precoding/combining schemes. We show that with multicell MMSE precoding/combining and a tiny amount of spatial channel correlation or large-scale fading variations over the array, the capacity increases without bound as the number of antennas increases, even under pilot contamination. More precisely, the result holds when the channel covariance matrices of the contaminating users are asymptotically linearly independent, which is generally the case. If also the diagonals of the covariance matrices are linearly independent, it is sufficient to know these diagonals (and not the full covariance matrices) to achieve an unlimited asymptotic capacity.
\end{abstract}

\begin{IEEEkeywords}
Massive MIMO, ergodic capacity, asymptotic analysis, spatial correlation, multi-cell MMSE processing, pilot contamination.\end{IEEEkeywords}

\section{Introduction} \label{sec-intro}

The Shannon capacity of a channel manifests the spectral efficiency (SE) that it supports. Massive MIMO (multiple-input multiple-output) improves the sum SE of cellular networks by spatial multiplexing of a large number of user equipments (UEs) per cell \cite{marzetta2010noncooperative}. It is therefore considered a key time-division duplex (TDD) technology for the next generation of cellular networks  \cite{Larsson2014,Andrews2014a,Larsson2017a}. The main difference between Massive MIMO and classical multiuser MIMO is the large number of antennas, $M$, at each base station (BS) whose signals are processed by individual radio-frequency chains. By exploiting channel estimates for coherent receive combining, the uplink signal power of a desired UE is reinforced by a factor $M$, while the power of the noise and independent interference does not increase. The same principle holds for the transmit precoding in the downlink. Since the channel estimates are obtained by uplink pilot signaling and the pilot resources are limited by the channel coherence time, the same pilots must be reused in multiple cells. This leads to pilot contamination which has two main consequences: the channel estimation quality is reduced due to pilot interference and the channel estimate of a desired UE is correlated with the channels to the interfering UEs that use the same pilot. Marzetta showed in his seminal paper \cite{marzetta2010noncooperative} that the interference from these UEs during data transmission is also reinforced by a factor $M$, under the assumptions of maximum ratio (MR) combining/precoding and independent and identically distributed (i.i.d.) Rayleigh fading channels. This means that pilot contamination creates a finite SE limit as $M \to \infty$.

The large-antenna limit has also been studied for other combining/precoding schemes, such as the minimum mean squared error (MMSE) scheme. Single-cell MMSE (S-MMSE) was considered in \cite{hoydis2013massive,Guo2014a,Krishnan2014a}, while multicell MMSE (M-MMSE) was considered in \cite{Ngo2012b,EmilEURASIP17}. The difference is that with M-MMSE, the BS makes use of estimates of the channels from the UEs in all cells, while with S-MMSE, the BS only uses channel estimates of the UEs in the own cell. In both cases, the SE was proved to have a finite limit as $M \to \infty$, under the assumption of i.i.d.~Rayleigh fading channels (i.e., no spatial correlation). 
In contrast, there are special cases of spatially correlated fading that give rise to rank-deficient covariance matrices \cite{Yin2013a,Adhikary2013,You2015a}. If the UEs that share a pilot have rank-deficient covariance matrices with orthogonal support, then pilot contamination vanishes and the SE can increase without bound. 
The covariance matrices $\vect{R}_1$ and $\vect{R}_2$ have orthogonal support if $\vect{R}_1 \vect{R}_2 = \vect{0}$. To understand this condition, note that for arbitrary covariance matrices 
\begin{equation} \label{eq:simple-covariance-matrices}
\vect{R}_1 = \begin{bmatrix} a & c \\ c^\star & b \end{bmatrix} \quad \vect{R}_2 = \begin{bmatrix} d & f \\ f^\star & e \end{bmatrix}
\end{equation}
every element of $\vect{R}_1 \vect{R}_2 $ must be zero. The first element is $ad+cf^\star$. If we model the practical covariance matrices of two randomly located UEs as realizations of a random variable with continuous distribution, then $ad+cf^\star=0$ occurs with zero probability.\footnote{For any continuous random variable $x$, the probability that $x$ takes a particular realization is zero, while the probability that $x$ takes a realization in a certain interval can be non-zero.
Hence, if $x =ad+cf^\star$ then $x=0$ occurs with zero probability.} 
Hence, orthogonal support is very unlikely in practice, although one can find special cases where it is satisfied.
The one-ring model for uniform linear arrays (ULAs) gives orthogonal support if the channels have non-overlapping angular support  \cite{Yin2013a,Adhikary2013,You2015a}, but the ULA microwave measurements in \cite{Gao2015a} show that the angular support of practical channels is highly irregular and does not lead to orthogonal support. In conclusion, practical covariance matrices do not have orthogonal support, at least not at microwave frequencies.

The literature contains several categories of methods for mitigation of pilot contamination, also known as \emph{pilot decontamination}.
The first category allocates pilots to the UEs in an attempt to find combinations where the covariance matrices have relatively different support \cite{Yin2013a,Adhikary2013,Li2013a,You2015a}. This method can substantially reduce pilot contamination, but can only remove the finite limit in the unlikely special case when the covariance matrices have orthogonal support.
The second category utilizes semi-blind estimation to separate the subspace of desired UE channels from the subspace of interfering channels \cite{Ngo2012a,Mueller2014b,Hu2016a,Yin2016a,Vinogradova2016a}. This method can fully remove pilot contamination if $M$ and the size of the channel coherence block go jointly to infinity \cite{Yin2016a}. Unfortunately, the channel coherence is fixed and finite in practice (this is why we cannot give unique pilots to every cell), thus we cannot approach this limit in practice. The third category uses multiple pilot phases with different pilot sequences to successively eliminate pilot contamination \cite{Zhang2014a,Vu2014a}, without the need for statistical information. However, the total pilot length is larger or equal to the total number of UEs, which would allow allocating mutually orthogonal pilots to all UEs and thus trivially avoiding the pilot contamination problem. This is not a scalable solution for networks with many cells. The fourth category is pilot contamination precoding that rejects interference by coherent joint transmission/reception over the entire network  \cite{ashikhmin2012pilot,Li2013b}. This method appears to achieve an unbounded SE, but this has not been formally proved and requires that the data for all UEs is available at every BS, which might not be feasible in practice.

In summary, it appears that pilot contamination is a fundamental issue that manifests a finite SE limit, except in unlikely special cases. We show in this paper that this is basically a misunderstanding, spurred by the popularity of analyzing suboptimal combining/precoding schemes, such as MR and S-MMSE, and focusing on unrealistic i.i.d.~Rayleigh fading channels (as in the prior work \cite{Ngo2012b,EmilEURASIP17} on M-MMSE). We prove that the SE increases without bound in the presence of pilot contamination when using M-MMSE combining/precoding, if the pilot-sharing UEs have asymptotically linearly independent covariance matrices. Note that $\vect{R}_1$ and $\vect{R}_2$ in \eqref{eq:simple-covariance-matrices} are linearly independent if $[a \, b \,c]^{\Ttran}$ and $[d \, e \,f]^{\Ttran}$ are non-parallel vectors, which happens almost surely for randomly generated covariance matrices. Hence, our results rely on a condition that is most likely satisfied in practice---it is the \emph{general case}, while prior works on the asymptotics of Massive MIMO have considered practically unlikely special cases. In contrast to prior work, no multicell cooperation is utilized herein and there is no need for orthogonal support of covariance matrices.
In the conference paper \cite{Bjornson2017a}, we proved the main result in a two-user uplink scenario.\footnote{After submitting our conference paper \cite{Bjornson2017a}, the related work \cite{Neumann2017a} appeared. That paper considers the mean squared error in the uplink data detection of a single cell with multiple UEs per pilot sequence. The authors show that the error goes asymptotically to zero when having linearly independent covariance matrices. However, the paper \cite{Neumann2017a} contains no mathematical analysis of the achievable SE.} In this paper, we prove the result for both uplink and downlink in a general setting. Section~\ref{section:two-user} proves and explains the intuition of the results in a two-user setup, while Section~\ref{section:multi-user} generalizes the results to a multicell setup. The results are demonstrated numerically in Section~\ref{sec:numerical-results} and the main conclusions are summarized in Section~\ref{section:conclusion}.

\subsubsection*{Notation} The Frobenius and spectral norms of a matrix $\vect{X}$ are denoted by $\| \vect{X} \|_F$ and $\| \vect{X} \|_2$, respectively. The superscripts $^{\Ttran}$, $^\star$ and $^{\Htran}$ denote transpose, conjugate, and Hermitian transpose, respectively. We use $\triangleq$ to denote definitions, whereas $\CN({\bf 0},{\bf R})$ denotes the circularly symmetric complex Gaussian distribution with zero mean and covariance matrix ${\bf R}$. The expected value of a random variable $x$ is denoted by $\mathbb{E}\{ x \}$ and the variance is denoted by $\mathbb{V}\{ x \}$. The $N \times N$ identity matrix is denoted by $\vect{I}_N$, while $\vect{0}_N$ is an $N \times N$ all-zero matrix and $\vect{1}_N$ is an $N \times 1$ all-one vector.
 We use $a_n \asymp b_n$ to denote $a_n -b_n \to_{n\to \infty}0$ (almost surely (a.s.)) for two (random) sequences $a_n$, $b_n$.

\section{Asymptotic Spectral Efficiency in a Two-User Scenario}
\label{section:two-user}

In this section, we prove and explain our main result in a two-user scenario, where a BS equipped with $M$ antennas communicates with UE~$1$ and UE~$2$ that are using the same pilot. This setup is sufficient to demonstrate why M-MMSE combining and precoding reject the coherent interference caused by pilot contamination. 
We consider a block-fading model where each channel takes one realization in a coherence block of $\tau_c$ channel uses and independent realizations across blocks.
We denote by ${\bf h}_{k} \in \mathbb{C}^{M}$ the channel from UE $k$ to the BS and consider Rayleigh fading with $\vect{h}_{k} \sim \CN \left( \vect{0}, \vect{R}_{k}  \right)$ for $k=1,2$, 
where ${\vect{R}_{k} \in \mathbb{C}^{M\times M}}$ with\footnote{This assumption implies that there is non-zero energy received from and transmitted to  each UE.} $\tr (\vect{R}_{k}  ) > 0$ is the channel covariance matrix, which is assumed to be known at the BS. The Gaussian distribution models the small-scale fading whereas the covariance matrix $\vect{R}_{k}$ describes the macroscopic effects. The normalized trace ${\beta_{k}= \frac{1}{M} \tr \left( \vect{R}_{k} \right)}$  determines the average large-scale fading between UE $k$ and the BS, while the eigenstructure of $\vect{R}_{k}$ describes the spatial channel correlation. A special case that is convenient for analysis is i.i.d.~Rayleigh fading with ${\vect{R}_{k} = \beta_{k}  \vect{I}_{M}}$ \cite{Marzetta2016a}, but it only arises in fully isotropic fading environments. In general, each covariance matrix has spatial correlation and large-scale fading variations over the array,
represented by non-zero off-diagonal elements and non-identical diagonal elements, respectively.

\subsection{Uplink Channel Estimation}
We assume that the BS and UEs are perfectly synchronized and operate according to a TDD 
protocol wherein the data transmission phase is preceded by an uplink pilot phase for channel estimation. Both UEs use the same $\taupu$-length pilot sequence $\bphiu \in \mathbb{C}^{\taupu}$ with elements such that $\| \bphiu \|^2  = \bphiu^{\Htran} \bphiu  = {1}$. The received uplink signal $\vect{Y}^{p} \in \mathbb{C}^{N\times \taupu}$ at the {BS} is given by
\begin{align}
\vect{Y}^{p}= \sqrt{\rho^{\rm{tr}}} \vect{h}_{1} \bphiu^{\Ttran} + \sqrt{\rho^{\rm{tr}}} \vect{h}_{2} \bphiu^{\Ttran} + \vect{N}^{p}
\end{align}
where $\rho^{\rm{tr}}$ is the normalized pilot power and $\vect{N}^{p} \in \mathbb{C}^{N\times \taupu}$ is the normalized receiver noise with all elements independently distributed as $\CN(0,1)$. The matrix $\vect{Y}^{p}$ is the observation that the {BS} utilizes to estimate ${\bf h}_1$ and ${\bf h}_2$. We assume that channel estimation is performed using the {MMSE} estimator given in the next lemma (the proof relies on standard estimation theory \cite{Kay_Book}).

\begin{lemma} \label{theorem:MMSE-estimate_h_jli}
The {MMSE} estimator of $\vect{h}_{k}$ for $k=1,2$, based on the observation $\vect{Y}^{p} $ at the {BS}, is
\begin{equation} \label{eq:MMSEestimator_h}
\begin{split}
\!\!\hat{\vect{h}}_{k}  =  \frac{1}{ \sqrt{\rho^{\rm{tr}}} }\vect{R}_{k}
 {\bf{Q}}^{-1} \vect{Y}^{p} \bphiu^{\star} 
\end{split}
\end{equation}
with ${\bf{Q}} = \frac{1}{\rho^{\rm{tr}}} \mathbb{E}\{ \vect{Y}^{p} \bphiu^{\star} ( \vect{Y}^{p} \bphiu^{\star}  )^{\Htran} \} = \vect{R}_{1} +  \vect{R}_{2} + \frac{1}{ \rho^{\rm{tr}}} \vect{I}_{M}$
being the normalized covariance matrix of the observation after correlating with the pilot sequence.
The estimate $\hat{\vect{h}}_{k} $ and the estimation error $\tilde{\vect{h}}_{k}= \vect{h}_{k} - \hat{\vect{h}}_{k}$ are independent random vectors distributed as $\hat{\vect{h}}_{k}  \sim \CN({\bf 0},\vect{\Phi}_{k})$ and $\tilde{\vect{h}}_{k}  \sim \CN({\bf 0},\vect{R}_{k} - \vect{\Phi}_{k})$ with $\vect{\Phi}_{k} = \vect{R}_{k}
{\bf{Q}}^{-1} \vect{R}_{k}$.
\end{lemma}

Interestingly, the estimates $\hat {\bf h}_1$ and $\hat {\bf h}_2$ are computed in an almost identical way in \eqref{eq:MMSEestimator_h}: the same matrix ${\bf{Q}}$ is inverted and multiplied with the same observation $\vect{Y}^{p} \bphiu^{\star}/\sqrt{\rho^{\rm{tr}}} $. 
The only difference is that for $\hat {\bf h}_k$ there is a multiplication with the UE's own channel covariance matrix $\vect{R}_{k}$ in \eqref{eq:MMSEestimator_h}, for $k=1,2$.
The channel estimates are thus correlated with correlation matrix
$
\vect{\Upsilon}_{12}   = {\mathbb{E}}\{\hat{\vect{h}}_{1}\hat{\vect{h}}_{2}^{\Htran}\}  = \vect{R}_{1}
{\bf{Q}}^{-1} \vect{R}_{2}$.
If $\vect{R}_{1}$ is invertible, then we can also write the relation between the estimates as $\hat{\vect{h}}_{2}    =  \vect{R}_{2} \vect{R}_{1}^{-1}  \hat{\vect{h}}_{1}$.
In the special case of i.i.d.~fading channels with $\vect{R}_{1}=\beta_{1} \vect{I}_{M}$ and $\vect{R}_{2}= \beta_{2} \vect{I}_{M}$, the two channel estimates are parallel vectors that only differ in scaling: $\hat{\vect{h}}_{2}    =  \frac{\beta_{2}}{\beta_{1}}  \hat{\vect{h}}_{1}$. This is an unwanted property caused by the inability of the {BS} to separate UEs that have transmitted the same pilot sequence over 
channels that are identically distributed (up to a scaling factor). In the alternative special case of $\vect{R}_{1} \vect{R}_{2}  = \vect{0}_M$, the two {UE} channels are located in orthogonal subspaces (i.e., have orthogonal support), which leads to zero correlation: $\vect{\Upsilon}_{12}= \vect{0}_M$. Consequently, it is theoretically possible to let two UEs share a pilot sequence without causing pilot contamination, if their covariance matrices satisfy the orthogonality condition $\vect{R}_{1} \vect{R}_{2}  = \vect{0}_M$. As described in Section~\ref{sec-intro}, none of these special cases occur in practice, therefore we will develop a general way to deal with the correlation of channel estimates caused by pilot contamination.

\subsection{Uplink Data Transmission}
During uplink data transmission, the received baseband signal at the BS is ${\bf y} \in \mathbb{C}^{M}$, given by
$
\vect{y}= \sqrt{\rho^{\rm {ul}}} \vect{h}_{1} s_1 + \sqrt{\rho^{\rm {ul}}} \vect{h}_{2} s_2 + \vect{n}
$,
where $s_k\sim\CN(0,1)$ is the information-bearing signal transmitted by UE~$k$, $\vect{n}\sim \CN(\vect{0},{\bf I}_M)$
is the independent receiver noise, and $\rho^{\rm {ul}}$ is the normalized transmit power. The BS detects the signal from UE $1$ by using a combining vector $\vect{v}_1\in \mathbb{C}^{M}$ to obtain $\vect{v}_1^{\Htran}\vect{y}$. Using a standard technique (see, e.g., \cite{hoydis2013massive,Marzetta2016a}), the ergodic uplink capacity of UE $1$ is lower bounded by 
\begin{align} \label{eq:SE-uplink-twousers}
\mathsf{SE}_{1}^{\rm {ul}} = \left( 1 - \frac{\taupu}{\tau_c}  \right) \mathbb{E} \left\{ \log_2  \left( 1 + \gamma_{1}^{\rm {ul}}  \right) \right\} \quad \textrm{[bit/s/Hz] }
\end{align}
where the expectation is with respect to the channel estimates. We refer to $\mathsf{SE}_{1}^{\rm {ul}}$ as an achievable SE.
The instantaneous effective signal-to-interference-and-noise ratio (SINR) $\gamma_{1}^{\rm {ul}}$ in \eqref{eq:SE-uplink-twousers} is

\begin{align} \nonumber
\gamma_{1}^{\rm {ul}}  &=  \frac{ |  \vect{v}_{1}^{\Htran} \hat{\vect{h}}_{1} |^2  }{{\mathbb{E}}\left\{ |  \vect{v}_{1}^{\Htran} \tilde{\vect{h}}_{1} |^2 + 
| \vect{v}_{1}^{\Htran} {\vect{h}}_{2} |^2
 + \frac{1}{\rho^{\rm {ul}}}\vect{v}_{1}^{\Htran}\vect{v}_{1}  
\Big| \hat{\bf{h}}_{1},\hat{\bf{h}}_{2}  \right\}}\\ &= \frac{ |  \vect{v}_{1}^{\Htran} \hat{\vect{h}}_{1} |^2  }{ 
 \vect{v}_{1}^{\Htran}  \left( \hat{\vect{h}}_{2} \hat{\vect{h}}_{2}^{\Htran} + \vect{Z}  \right) \vect{v}_{1}  } \label{eq:gamma1}
\end{align}
with
$
\vect{Z} =  \sum_{k=1}^{2} (\vect{R}_{k} - \vect{\Phi}_{k}) + \frac{1}{\rho^{\rm {ul}}}  \vect{I}_M.
$
Since $\gamma_{1}^{\rm{ul}}$ is a generalized Rayleigh quotient, the SINR is maximized by  \cite{Ngo2012b,EmilEURASIP17} 
\begin{align} \label{v_k_MMSE}
\vect{v}_1= \left( \sum_{k=1}^2\hat{\vect{h}}_{k} \hat{\vect{h}}_{k}^{\Htran} + \vect{Z}  \right)^{-1} \hat{\vect{h}}_{1}.
\end{align}
This is called MMSE combining since \eqref{v_k_MMSE} not only maximizes the instantaneous {SINR} $\gamma_{1}^{\rm {ul}}$, but also minimizes  ${\mathbb{E}}\{ |x_{1} - \vect{v}_{1}^{\Htran} {\bf y}   |^2  \, |{{\hat{\bf h}}_{1}},{{\hat{\bf h}}_{2}}\}$ which is the mean squared error (MSE) in the data detection (conditioned on the channel estimates). Plugging \eqref{v_k_MMSE} into \eqref{eq:gamma1} yields
\begin{align} \label{eq:gamma1_MMSE}
\gamma_{1}^{\rm {ul}}  &=  \hat{\vect{h}}_{1}^{\Htran}\left(  \hat{\vect{h}}_{2} \hat{\vect{h}}_{2}^{\Htran} + {\bf Z}\right)^{-1} \hat{\vect{h}}_{1}.\end{align}
We will now analyze the asymptotic behavior of $\mathsf{SE}_{1}^{\rm {ul}}$ and $\gamma_{1}^{\rm{ul}}$ as $M\to \infty$. To this end, we make the following technical assumptions:

\begin{assumption}\label{assumption_1} For $k=1,2$,  $\mathop {\liminf}\limits_M\frac{1}{{M}}\tr ( \vect{R}_{k} ) > 0$ and $ \mathop {\limsup}\limits_M \| \vect{R}_{k}\|_2 < \infty$.

\end{assumption}
\begin{assumption}\label{assumption_2} For $\boldsymbol{\lambda} = [\lambda_1, \lambda_2]^{\Ttran} \in \mathbb{R}^2$ and $i=1,2$, 
\begin{align} \label{eq:assumption_2_relaxed}
\mathop {\liminf}\limits_M \inf_{\{\boldsymbol{\lambda}: \, \lambda_i=1\}}  \frac{1}{{M}}  \left\| \lambda_1 \vect{R}_{1} + \lambda_2 \vect{R}_{2} \right\|_F^2 > 0.
\end{align}
\end{assumption} 

The first assumption is a well established way to model that the array gathers more energy as $M$ increases and also that this energy originates from many spatial dimensions \cite{hoydis2013massive}. In particular, it is a sufficient condition for asymptotic channel hardening; that is, $\| \vect{h}_{k} \|^2/ \mathbb{E}\{ \| \vect{h}_{k} \|^2 \} \to 1$  in probability as $M \to \infty$. 
The second assumption requires $\vect{R}_1$ and $\vect{R}_2$ to be \emph{asymptotically} linearly independent, in the sense that if one of the matrices is scaled to resemble the other one, the subspace in which the matrices differ has an energy proportional to $M$. Note that this is a stronger condition than linear independence, defined as $\inf_{\{\boldsymbol{\lambda}: \, \lambda_i=1\}}   \| \lambda_1 \vect{R}_{1} + \lambda_2 \vect{R}_{2} \|_F^2 > 0$ for $i=1,2$, which is satisfied even if the matrices only differ in one element. We will elaborate further on Assumption~\ref{assumption_2} in Section~\ref{subsec:interpretation}.

The following is the first of the main results of this paper:

\begin{theorem} \label{theorem:MMSE}
If MMSE combining is used, then under Assumptions \ref{assumption_1} and \ref{assumption_2}, the instantaneous effective SINR $\gamma_{1}^{\rm{ul}} $ increases a.s.~unboundedly as $M\to \infty$. Hence, $\mathsf{SE}_{1}^{\rm {ul}}$ increases unboundedly as $M\to \infty$.
\end{theorem}
\begin{IEEEproof}
The proof is given in Appendix~B.
\end{IEEEproof}

\begin{remark}\label{rem:2-UE-UL}
From the proof in Appendix~B, we can see that $\gamma_{1}^{\rm{ul}}/M$ has a non-zero asymptotic limit, which implies that the SE grows towards infinity as $\log_2(M)$. While Theorem~\ref{theorem:MMSE} only considers UE~1, one only needs to interchange the UE indices to prove that the SE of UE~2 also grows unboundedly as $M\to \infty$. Hence, an unlimited asymptotic SE is simultaneously achievable for both UEs. Since the SE is a lower bound on  capacity, we conclude that the asymptotic capacity is also unlimited.
\end{remark}

Observe that if ${\bf R}_1$ and ${\bf R}_2$ are linearly dependent, i.e., $\vect{R}_1 = \eta \vect{R}_2$, then Assumption~\ref{assumption_2} does not hold. Under these circumstances, $\hat {\vect{h}}_{2} = \frac{1}{\eta}\hat{\vect{h}}_{1}$ and by applying Lemma \ref{MIL} in Appendix A we obtain \begin{align} \label{eq:gamma1_MMSE_linearly_dependent}
\gamma_{1}^{\rm {ul}}  &=  \frac{\hat{\vect{h}}_{1}^{\Htran}{\bf Z}^{-1}\hat{\vect{h}}_{1}}{1 + \frac{1}{\eta^2}\hat{\vect{h}}_{1}^{\Htran}{\bf Z}^{-1}\hat{\vect{h}}_{1}}\end{align}
from which, it is straightforward to show that $\gamma_1^{\rm {ul}} \asymp \eta^2$ (by dividing and multiplying each term by $M$ and using Lemma~\ref{lemma3} in Appendix A). This implies that $\mathsf{SE}_1^{\rm {ul}} $ converges to a finite quantity when $M\to \infty$, as Marzetta showed in his seminal paper  \cite{marzetta2010noncooperative} for the special case of ${\bf R}_1=\eta {\bf R}_2={\bf I}_M$.

\subsection{Downlink Data Transmission}
During the downlink data transmission, the BS transmits the signal ${\bf x} \in \mathbb{C}^M$. This signal is given by ${\bf x} = \sqrt{\rho^{\rm {dl}} }{\bf w}_1\varsigma_1 + \sqrt{\rho^{\rm {dl}}}{\bf w}_2\varsigma_2$, where $\varsigma_k\sim\CN(0,1)$ is the information-bearing signal transmitted to UE~$k$, $\rho^{\rm {dl}}$ is the normalized downlink transmit power, and ${\bf w}_k$ is the precoding vector associated with UE $k$. This precoding vector  satisfies $\mathbb{E} \left\{\|{\bf w}_k\|^2\right\} =1$, so that $\mathbb{E} \left\{\|{\bf w}_k\varsigma_k\|^2\right\} =\rho^{\rm{dl}}$ is the downlink transmit power allocated to UE $k$. The received downlink signal $z_1$ at UE~1 is\footnote{For notational convenience, we treat $\vect{h}_{1}^{\Htran}$ and $\vect{h}_{2}^{\Htran} $ as the downlink channels, instead of $\vect{h}_{1}^{\Ttran}$ and $\vect{h}_{2}^{\Ttran}$. This has no impact on the SE since the difference is only in a complex conjugate.}
\begin{align}\notag
z_1 &=  \sqrt{\rho^{\rm {dl}}}\vect{h}_{1}^{\Htran} {\vect{w}}_{1} \varsigma_1 + \sqrt{\rho^{\rm {dl}}}\vect{h}_{1}^{\Htran} {\vect{w}}_{2} \varsigma_2+ n_1
\\\notag
&=\sqrt{\rho^{\rm {dl}}}{\mathbb{E}}\left\{\vect{h}_{1}^{\Htran} {\vect{w}}_{1}\right\}\varsigma_1 + \sqrt{\rho^{\rm {dl}}}(\vect{h}_{1}^{\Htran} {\vect{w}}_{1} - {\mathbb{E}}\left\{\vect{h}_{1}^{\Htran} {\vect{w}}_{1}\right\})\varsigma_1 \\&\hspace{3.1cm}+ \sqrt{\rho^{\rm {dl}}} \vect{h}_{1}^{\Htran} {\vect{w}}_{2} \varsigma_2+ n_1\label{eq:received_signal_DL}
\end{align}
where $n_1\sim \CN(0,1)$ is the normalized receiver noise.
The first term in \eqref{eq:received_signal_DL} is the desired signal received over the deterministic average precoded channel ${\mathbb{E}}\left\{\vect{h}_{1}^{\Htran} {\vect{w}}_{1}\right\}$, while the remaining 
terms are random variables with unknown realizations. By treating these terms as noise in the signal detection \cite{hoydis2013massive,Marzetta2016a}, the downlink ergodic channel capacity of UE 1 can be lower bounded by 
\begin{align} \label{eq:SE-downlink-twousers}
\mathsf{SE}_{1}^{\rm dl} = \left( 1 - \frac{\taupu}{\tau_c}  \right) \log_2  \left( 1 + \gamma_{1}^{\rm dl}  \right) \quad \textrm{[bit/s/Hz] }
\end{align}
with the effective SINR
\begin{align} \label{eq:gamma1_DL}
\gamma_{1}^{\rm {dl}}  
= \frac{ |  {\mathbb{E}}\{\vect{h}_{1}^{\Htran} {\vect{w}}_{1} \} |^2  }{ {\mathbb{E}}\left\{| \vect{h}_{1}^{\Htran} {\vect{w}}_{2}|^2\right\} + {\mathbb{V}}\left\{ \vect{h}_{1}^{\Htran} {\vect{w}}_{1}\right\} + \frac{1}{\rho^{\rm dl}}}.
\end{align}
Since UE~$1$ only needs to know ${\mathbb{E}}\left\{\vect{h}_{1}^{\Htran} {\vect{w}}_{1}\right\}$ and the total variance of the second to fourth term in \eqref{eq:received_signal_DL}, the SE in \eqref{eq:SE-downlink-twousers} is achievable in the absence of downlink channel estimation.
In contrast to the uplink, there is no precoding that is always optimal \cite{Bjornson2014d}. However, motivated by uplink-downlink duality \cite{EmilEURASIP17}, a reasonable suboptimal choice is the so-called MMSE precoding
\begin{align}\label{eq:Section3_precoding}
{\bf w}_k = \frac{\vect{v}_k}{\sqrt{{\mathbb{E}}\left\{\|\vect{v}_k\|^2\right\}}} = \sqrt{\vartheta_k} \left( \sum_{k=1}^2\hat{\vect{h}}_{k} \hat{\vect{h}}_{k}^{\Htran} + \vect{Z}  \right)^{-1} \hat{\vect{h}}_{k}
\end{align}
where $\vect{v}_k = ( \sum_{k=1}^2\hat{\vect{h}}_{k} \hat{\vect{h}}_{k}^{\Htran} + \vect{Z}  )^{-1} \hat{\vect{h}}_{k}$ is MMSE combining and $\vartheta_k = ({\mathbb{E}}\left\{\|\vect{v}_k\|^2\right\})^{-1}$ is a scaling factor. The following is the second main result of this paper:
\begin{theorem} \label{theorem:MMSE_precoding}
If MMSE precoding is used, then under Assumptions \ref{assumption_1} and \ref{assumption_2} the effective SINR $\gamma_{1}^{\rm{dl}} $ increases unboundedly as $M\to \infty$. Hence, $\mathsf{SE}_{1}^{\rm {dl}}$ increases unboundedly as $M\to \infty$.
\end{theorem}
\begin{IEEEproof}
The proof is given in Appendix~D.
\end{IEEEproof}

This theorem shows that, under the same conditions as in the uplink, the downlink SE (and thus the capacity) increases without bound as $M\to \infty$. The asymptotic SE growth is proportional to $\log_2(M)$, since the proof in Appendix~D shows that $\gamma_{1}^{\rm{dl}}/M$ has a non-zero asymptotic limit. UE~2 can simultaneously achieve an unbounded SE, which is proved directly by interchanging the UE indices.

\subsection{Interpretation and Generality}
\label{subsec:interpretation}

\begin{figure}[t!]
\begin{center} 
\includegraphics[width=.45\columnwidth]{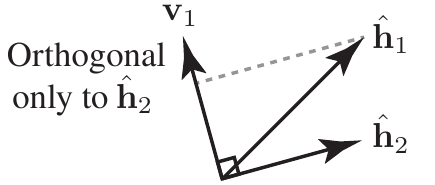}
\end{center}
\caption{If the pilot-contaminated channel estimates are linearly independent (i.e., not parallel), there exists a combining vector $\vect{v}_1$ that rejects the pilot-contaminated interference from UE~2 in the uplink, while the desired signal remains due to $\vect{v}_1^{\Htran} \hat{\vect{h}}_1 \neq 0$.
Similarly, if $\vect{w}_1 = \vect{v}_1/\sqrt{\mathbb{E}\{ \|  \vect{v}_1 \|^2 \}}$ is used as precoding vector, then no pilot-contaminated coherent interference is caused to UE~2 in the downlink.} \label{figureOrthogonality} 
\end{figure}

Theorems~\ref{theorem:MMSE} and \ref{theorem:MMSE_precoding} show that the SE (and thus the capacity) under pilot contamination is asymptotically unlimited if Assumption~\ref{assumption_2} holds.
To gain an intuitive interpretation of this underlying assumption, recall from \eqref{eq:MMSEestimator_h} that $\hat{\vect{h}}_1 = \vect{R}_{1} \vect{a}$ and $\hat{\vect{h}}_2 = \vect{R}_{2} \vect{a}$, where $\vect{a}= \frac{1}{ \sqrt{\rho^{\rm{tr}}} } {\bf{Q}}^{-1} \vect{Y}^{p} \bphiu^{*} $ is the same for both UEs.
Hence, $\hat{\vect{h}}_1$ and $\hat{\vect{h}}_2$ are (asymptotically) linearly independent when $\vect{R}_{1}$ and $\vect{R}_{2}$ are (asymptotically) linearly independent, except for special choices of $\vect{a}$. As illustrated in Fig.~\ref{figureOrthogonality}, it is then possible to find a combining vector $\vect{v}_1$ (or precoding vector $\vect{w}_1$) that is orthogonal to $\hat{\vect{h}}_2$, while being non-orthogonal to $\hat{\vect{h}}_1$. 
Similarly, one can find $\vect{v}_2$ (and $\vect{w}_2$) such that $\vect{v}_2^{\Htran}  \hat{\vect{h}}_1 = 0$ and  $\vect{v}_2^{\Htran} \hat{\vect{h}}_2 \neq 0$.
For example, if we define $\hat{\vect{H}} = [  \hat{\vect{h}}_1 \,\, \hat{\vect{h}}_2 ] \in \mathbb{C}^{M \times 2}$, then the zero-forcing (ZF) combining vectors
\begin{equation} \label{eq:basic-ZF}
\big[ \vect{v}_1 \,\, \vect{v}_2 \big]= \hat{\vect{H}} \left( \hat{\vect{H}}^{\Htran} \hat{\vect{H}} \right)^{-1}
\end{equation}
satisfy these conditions. Note that $\hat{\vect{H}}^{\Htran} \hat{\vect{H}}$ is only invertible if the channel estimates (columns in  $\hat{\vect{H}}$) are linearly independent.
Using ZF as defined in \eqref{eq:basic-ZF}, we get $\vect{v}_1^{\Htran}  \hat{\vect{h}}_2 = 0$ and  $\vect{v}_1^{\Htran} \hat{\vect{h}}_1 = 1$. If the channel estimates are also asymptotically linearly independent, it follows\footnote{Notice that, by applying Lemma 3 in Appendix A, we have $\frac{1}{M}[\hat{\vect{H}}^{\Htran} \hat{\vect{H}}]_{nm} \asymp \frac{1}{M}\tr(\vect{R}_{n}
{\bf{Q}}^{-1} \vect{R}_{m})$. If the channel estimates are asymptotically linearly independent, then $\frac{1}{M}\hat{\vect{H}}^{\Htran} \hat{\vect{H}}$ is invertible as $M\to \infty$ and thus $\| \vect{v}_1 \|^2 = \frac{1}{M}\big[(\frac{1}{M}\hat{\vect{H}}^{\Htran} \hat{\vect{H}})^{-1}]_{11}\asymp 0$.} that $ \| \vect{v}_1 \|^2 \to 0$ as $M \to \infty$; that is, we can reject the coherent interference and get unit signal gain, while at the same time using the array gain to make the noise term
$\frac{1}{\rho^{\rm {ul}}}\vect{v}_{1}^{\Htran}\vect{v}_{1}   = \frac{1}{\rho^{\rm {ul}}} \| \vect{v}_1 \|^2$
vanish asymptotically. Since optimal MMSE combining (and also MMSE precoding) provides a higher SINR than the heuristic ZF scheme in \eqref{eq:basic-ZF}, it also rejects the coherent interference while retaining an array gain that grows with~$M$.

To further explain the implications of Assumption~\ref{assumption_2}, we provide the following three examples.

\begin{example} \label{example1}
Consider a two-user scenario with
\begin{equation}
 \vect{R}_1 = \begin{bmatrix} 2 \vect{I}_N & \vect{0} \\ \vect{0} & \vect{I}_{M-N} \end{bmatrix} \qquad \vect{R}_2 = \vect{I}_M
 \end{equation}
 where the covariance matrices have full rank and are only different in the first $N$ dimensions.
For any given $M$, we notice that the argument of \eqref{eq:assumption_2_relaxed} for UE~$i=1$ becomes
 \begin{align}\notag
& \inf_{\lambda_2} \frac{1}{{M}} \| \vect{R}_{1} + \lambda_2 \vect{R}_{2} \|_F^2 \\&=  \inf_{\lambda_2} \frac{N(2+\lambda_2)^2+(M-N)(1+\lambda_2)^2}{M} =\frac{(M-N)N}{M^2} \label{eq:example1}
 \end{align}
 where the infimum is attained by $\lambda_2 = -(M+N)/M$. Note that \eqref{eq:example1} goes to zero as $M \to \infty$ if $N$ is constant, while it has the non-zero limit $(1-\alpha)\alpha$ if $N = \alpha M$, for some $0 < \alpha < 1$. In the latter case, the matrices $ \{ \vect{R}_1, \vect{R}_2 \}$ satisfy \eqref{eq:assumption_2_relaxed}. Interestingly, although the covariance matrices are diagonal, they are still asymptotically linearly independent and the subspace in which they differ has rank $\min(N,M-N)= M \min(\alpha, (1-\alpha) ) $, which is proportional to $M$.
 
Let us further exemplify the interference rejection by considering ZF combining, which provides lower SINR than MMSE combining, but gives more intuitive expressions. Assume for the sake of simplicity that the channel realizations are such that $\frac{1}{ \sqrt{\rho^{\rm{tr}}} } {\bf{Q}}^{-1} \vect{Y}^{p} \bphiu^{*}  = \vect{1}_M$, which gives 
$\hat{\vect{h}}_1 = \vect{R}_{1} \vect{1}_M = [2 \vect{1}_N^{\Ttran} \, \, \vect{1}_{M-N}^{\Ttran}]^{\Ttran}$ and $\hat{\vect{h}}_2 = \vect{R}_{2} \vect{1}_M = \vect{1}_M$. 
The ZF combining vectors are then given by
\begin{equation}
\big[ \vect{v}_1 \,\, \vect{v}_2 \big]= \hat{\vect{H}} \left( \hat{\vect{H}}^{\Htran} \hat{\vect{H}} \right)^{-1} \!\!\!\!\!\!= 
\begin{bmatrix} \frac{1}{N} \vect{1}_N & -\frac{1}{N} \vect{1}_{N} \\  -\frac{1}{M-N} \vect{1}_{M-N}&  \frac{2}{M-N} \vect{1}_{M-N} \end{bmatrix}.
\end{equation}
If we set $\rho^{\rm {ul}}=\rho^{\rm {ul}}=1$ for simplicity, the instantaneous effective SINR in \eqref{eq:gamma1} for UE~1  becomes
\begin{align} \notag
\gamma_{1}^{\rm {ul}}  
&= \frac{ |  \vect{v}_{1}^{\Htran} \hat{\vect{h}}_{1} |^2  }{ 
|  \vect{v}_{1}^{\Htran} \hat{\vect{h}}_{2} |^2 + \sum_{k=1}^{2}  \vect{v}_{1}^{\Htran}(\vect{R}_{k} - \vect{\Phi}_{k})  \vect{v}_{1} +  \| \vect{v}_{1}\|^2}
 \\&= \frac{ 1  }{ 0 + \frac{7}{4N} + \frac{4}{3(M-N)}    + \frac{M}{N(M-N)} } 
\end{align}
where the coherent interference from UE~2 is zero. The remaining terms go asymptotically to zero if $N = \alpha M$, for  $0 < \alpha < 1$, in which case $\gamma_{1}^{\rm {ul}}$ grows without bound, as expected from Theorem~\ref{theorem:MMSE}.
 \end{example}

In the second example, we consider a scenario where Assumption~\ref{assumption_2} is not satisfied.
 
 \begin{example} \label{example2}
Channels with i.i.d.~fading, where the covariance matrices are $\vect{R}_1 = \beta_1 {\bf I}_M$ and $\vect{R}_2 = \beta_2 {\bf I}_M$, are a notable case when the covariance matrices are not linearly independent. However, any such case is non-robust to perturbations of the matrix elements. Suppose we replace $\vect{R}_1$ with
\begin{equation} \label{eq:matrix-example-independence2}
\vect{R}_1 = \beta_1 \left[ {\begin{array}{*{20}{c}}
{{\epsilon _1}}&0& \cdots \\
0& \ddots &0\\
 \vdots &0&{{\epsilon _M}}
\end{array}} \right]
\end{equation}
where $\epsilon_{1},\ldots,\epsilon_{M}$ are i.i.d.~positive random variables. This modeling is motivated by the measurement results in \cite{Gao2015b}, which shows that there are a few dB of large-scale fading variations over the antennas in a ULA.  For UE~$i=1$, we have
 \begin{align} \notag
\mathop {\liminf}\limits_M &\inf_{\lambda_2} \frac{1}{{M}} \| \vect{R}_{1} + \lambda_2 \vect{R}_{2} \|_F^2 \\\notag&= \mathop {\liminf}\limits_M \inf_{\lambda_2} \frac{1}{M}\sum_{m=1}^{M} (\beta_1 \epsilon_m+\lambda_2\beta_2)^2   \\ & \mathop {=}^{(a)}  \mathop {\liminf}\limits_M \beta_1^2\frac{1}{M}\sum_{m=1}^{M} {\left( \epsilon_m-\frac{1}{M}\sum\limits_{n=1}^{M}\epsilon_n\right)^2} \notag\\&\mathop {=}^{(b)} \beta_1^2 \mathbb{E}\{ (\epsilon_m-\mathbb{E}\{ \epsilon_m \})^2\}
 \end{align}
where $(a)$ is obtained from the fact that $\lambda_2 = -\frac{\beta_1}{\beta_2}\frac{1}{M}\sum_{n=1}^{M} {\epsilon_n}$ minimizes $\frac{1}{M}\sum_{m=1}^{M} (\beta_1 \epsilon_m+\lambda_2\beta_2)^2$ and $(b)$ follows from the strong law of large numbers. Note that $\mathbb{E}\{ (\epsilon_m-\mathbb{E}\{ \epsilon_m \})^2\}$ in the last expression is the variance of $ \epsilon_m$. Since every random variable has non-zero variance and $\beta_1>0$, we conclude that $ \{ \vect{R}_1, \vect{R}_2 \}$ satisfy \eqref{eq:assumption_2_relaxed} and thus Assumption~\ref{assumption_2} holds.
\end{example}
 
The key implication from Example~\ref{example2} is that all cases where $\vect{R}_1$ and $\vect{R}_2$ are equal (up to a scaling factor) are non-robust to random perturbations and thus anomalies. Since practical propagation environments are irregular and behave randomly (see the measurements reported in \cite{Gao2015a,Gao2015b}), linearly dependent covariance matrices are not appearing in practice and Assumption~\ref{assumption_2} is generally satisfied. In other words, it is fair to say that the uplink and downlink SEs grow without bound as $M \to \infty$ in general, while the special cases when it does not occur are of no practical importance. We end this subsection with a comparison with related work and a remark regarding acquisition of channel statistics.

\begin{example}[Comparison with \cite{ashikhmin2012pilot,Li2013b}] \label{examplenew}
Consider a BS with two distributed arrays of $M'=M/2$ antennas that serve two UEs having the covariance matrices
\begin{equation}
 \vect{R}_1 = \begin{bmatrix} b_{11} \vect{I}_{M'} & \vect{0} \\ \vect{0} & b_{12} \vect{I}_{M'} \end{bmatrix} \quad \vect{R}_2 =  \begin{bmatrix} b_{21} \vect{I}_{M'} & \vect{0} \\ \vect{0} & b_{22} \vect{I}_{M'} \end{bmatrix}
 \end{equation}
 with $b_{11}, b_{12}, b_{21}, b_{22}>0$. These covariance matrices are (asymptotically) linearly independent if $b_{11} b_{22} \neq b_{12} b_{21}$, in which case the uplink and downlink SEs grow without bound with MMSE or ZF.
 
The exemplified setup is equivalent to the multicell joint transmission scenario considered in the pilot contamination precoding works \cite{ashikhmin2012pilot,Li2013b} in which the heuristic vectors
\begin{equation} \label{eq:PCP}
\big[ \vect{v}_1 \,\, \vect{v}_2 \big]= \begin{bmatrix} \frac{1}{M'} \vect{y}^{p}_1 & \vect{0} \\  \vect{0} &  \frac{1}{M'} \vect{y}^{p}_2  \end{bmatrix}
\begin{bmatrix} b_{11} & b_{12} \\  b_{21} &  b_{22}  \end{bmatrix}^{-1}
\end{equation}
are used for combining and precoding, and $\vect{y}^{p}_1,\vect{y}^{p}_2 \in \mathbb{C}^{M'}$ are obtained from the received pilot signals as
$ [ (\vect{y}^{p}_1)^{\Ttran} \, (\vect{y}^{p}_2)^{\Ttran} ]^{\Ttran} =  \vect{Y}^{p} \bphiu^{\star}/\rho^{\rm{tr}}$. These vectors are specifically designed to make $\big[ \vect{h}_1 \,\, \vect{h}_2 \big]^{\Htran}\big[ \vect{v}_1 \,\, \vect{v}_2 \big]\asymp {\bf I}_2$
as $M \to \infty$, and thus this method has the same asymptotic behavior as ZF in the special case of block-diagonal covariance matrices where each block is a scaled identity matrix. Note that the matrix inverse in \eqref{eq:PCP} only exists if $b_{11} b_{22} \neq b_{12} b_{21}$, which is again the condition for linear independence of the covariance matrices. Since pilot contamination precoding can only be applied in special multicell cooperation cases, MMSE combining/precoding is generally the preferable choice.
\end{example}

\begin{remark}[Acquiring Covariance Matrices]
Theorems~\ref{theorem:MMSE} and \ref{theorem:MMSE_precoding} exploit the MMSE estimator and thus the BS needs to know the (deterministic) channel statistics. In particular, the BS can only compute the MMSE estimate $\hat {\bf h}_k$ in Lemma~\ref{theorem:MMSE-estimate_h_jli} if it knows $\vect{R}_{k}$ and also the sum  $\vect{R}_{1}+\vect{R}_{2}$ of the two covariance matrices. In practice, $\vect{R}_{k}$ can be estimated by a regularized sample covariance matrix, given realizations of $\vect{h}_{k}$ over multiple resource blocks (e.g., different times and frequencies) where this channel is either observed in only noise \cite{Yin2013a,Shariati2014a,Sun2015a} or where some observations are regular pilot transmissions containing the desired channel plus interference/noise and some contain only the interference/noise \cite{Bjornson2016c}. It seems that around $M$ samples are needed to obtain a sufficiently accurate covariance estimate \cite{Bjornson2016c}.
The covariance estimation can be further improved if the channels have a known structure.
 For example, \cite{Haghighatshoar2017a} provides algorithms for estimating the covariance matrices of channels that have limited angle-delay support that is also separable between users.
\end{remark}

\subsection{Achievable SE with Partial Knowledge of Covariance Matrices}\label{sec:approximate_MMSE_two_user}

If the BS does not have full knowledge of the covariance matrices, an alternative method for channel estimation is to estimate each entry of $\vect{h}_{k}$ separately, ignoring the correlation among the elements. This leads to the element-wise MMSE (EW-MMSE) estimator (called diagonalized estimator in \cite{Shariati2014a}) that utilizes only the main diagonals of $\vect{R}_{1}$ and $\vect{R}_{2}$. The diagonals can be estimated efficiently using a small number of samples, that does not need to grow with $M$ \cite{Bjornson2016c,Shariati2014a}.
\begin{lemma}\label{lemma:EW-MMSE}
Based on the observation $[ \vect{Y}^{p} \bphiu^{*} ]_{i}$, the BS can compute the EW-MMSE estimate of the $i$th element of $\vect{h}_{k}$ as 
\begin{align}
[\hat{\vect{h}}_{k}]_i  = \frac{1}{ \sqrt{\rho^{\rm{tr}}} }\frac{ [ \vect{R}_{k} ]_{ii}}{ [ \vect{R}_{1} ]_{ii} + [ \vect{R}_{2} ]_{ii}  +  \frac{1}{ {\rho^{\rm{tr}}} }  }[ \vect{Y}^{p} \bphiu^{*}]_{i}.
\end{align}
\end{lemma}
We may write $\hat{\vect{h}}_{k} $ in Lemma~\ref{lemma:EW-MMSE} in matrix form as
\begin{equation}\label{eq:G_101}
\hat{\vect{h}}_{k}  =  \frac{1}{ \sqrt{\rho^{\rm{tr}}} } {\bf D}_k \boldsymbol{\Lambda}^{-1} \vect{Y}^{p} \bphiu^{*}
\end{equation}
where ${\bf D}_k \in \mathbb{R}^{M\times M}$ and $\boldsymbol{\Lambda}\in \mathbb{R}^{M\times M}$ are diagonal matrices with elements $\{[ \vect{R}_{k} ]_{ii}: i=1,\ldots,M\}$ and $\{[ \vect{R}_{1} ]_{ii} + [ \vect{R}_{2} ]_{ii}  +  \frac{1}{ {\rho^{\rm{tr}}} } : i=1,\ldots,M\}$, respectively. {Notice that Assumption~\ref{assumption_1} implies that\footnote{This easily follows by observing that $\tr({\bf R}_k) = \tr({\bf D}_k)$ and also that $[{\bf D}_k]_{ii} = [{\bf R}_k]_{ii} \le \left\|{\bf R}_k\right\|_2$ since ${\bf R}_k$ is Hermitian.} $ \liminf_M \frac{1}{{M}}\tr ( \vect{D}_{k} ) > 0$ and $ \limsup_M \| \vect{D}_{k}\|_2 < \infty$ for $k=1,2$}.
To quantify the achievable SE when using EW-MMSE, similar to the downlink we exploit the use-and-then-forget SE bound \cite{Marzetta2016a}, which is less tight than \eqref{eq:SE-uplink-twousers} but does not require the use of MMSE channel estimation. The uplink ergodic capacity of UE 1 can be thus lower bounded by ${\underline{\mathsf{SE}}}_{1}^{\rm ul} = ( 1 - \frac{\taupu}{\tau_c}  )  \log_2  ( 1 + {\underline{\gamma}}_{1}^{\rm ul}  )$ [bit/s/Hz] with 
\begin{align} \label{eq:G_102}
{\underline{\gamma}}_{1}^{\rm ul}
= \frac{ |  {\mathbb{E}}\{\vect{v}_{1}^{\Htran} {\vect{h}}_{1}\} |^2  }{ {\mathbb{E}}\left\{| \vect{v}_{1}^{\Htran} {\vect{h}}_{2}|^2\right\} + {\mathbb{V}}\{ \vect{v}_{1}^{\Htran} {\vect{h}}_{1} \} + \frac{1}{\rho^{\rm ul}}  {\mathbb{E}}\left\{\|\vect{v}_{1} \|^2\right\} }.
\end{align}
This bound is valid for any channel estimation and any combining scheme. A reasonable choice for $\vect{v}_1$ is the approximate MMSE combining vector:
\begin{align} \label{eq:G_104}
\vect{v}_1= \bigg( \sum_{k=1}^2\hat{\vect{h}}_{k} \hat{\vect{h}}_{k}^{\Htran} + \vect{S}\bigg)^{-1} \hat{\vect{h}}_{1}
\end{align}
where $\hat{\vect{h}}_{1},\hat{\vect{h}}_{2}$ are computed as in \eqref{eq:G_101} and
$\vect{S}$ is diagonal and given by $\vect{S} =  \sum_{k=1}^{2} \Big(\vect{D}_{k} - \vect{D}_{k}\boldsymbol{\Lambda}^{-1}\vect{D}_{k}\Big) + \frac{1}{\rho^{\rm {ul}}}  \vect{I}_M$. Note that \eqref{eq:G_104} is equivalent to the MMSE combining in \eqref{v_k_MMSE} when the covariance matrices are diagonal.
 We will now analyze how ${\underline{\gamma}}_{1}^{\rm ul}$ behaves asymptotically as $M\to \infty$ when $\vect{v}_1$ is given by \eqref{eq:G_104}. To this end, we impose the following assumption, which states that $\vect{D}_1$ and $\vect{D}_2$ are asymptotically linearly independent (i.e., the diagonals of $\vect{R}_1$ and $\vect{R}_2$ are asymptotically linearly independent).
\begin{assumption}\label{assumption_6}For $\boldsymbol{\lambda} = [\lambda_1, \lambda_2]^{\Ttran} \in \mathbb{R}^2$ and $i=1,2$,  \begin{align} \label{eq:assumption_6}
\mathop {\liminf}\limits_M \inf_{\{\boldsymbol{\lambda}: \, \lambda_i=1\}}  \frac{1}{{M}}  \left\| \lambda_1 \vect{D}_{1} + \lambda_2 \vect{D}_{2} \right\|_F^2 > 0.
\end{align}\end{assumption} 
The following is the third main result of this paper:

\begin{theorem} \label{theorem:EW-MMSE_precoding}
If ${\bf v}_1$ in \eqref{eq:G_104} is used with $\hat{\vect{h}}_{1},\hat{\vect{h}}_{2}$ given by \eqref{eq:G_101}, then under Assumptions~\ref{assumption_1} and \ref{assumption_6}, the SINR $\underline\gamma_{1}^{\rm{ul}} $ increases unboundedly as $M\to \infty$. Hence, ${\underline{\mathsf{SE}}}_{1}^{\rm ul}$ increases unboundedly as $M\to \infty$.
\end{theorem}
\begin{IEEEproof}
The proof is given in Appendix~E.
\end{IEEEproof}

As a consequence of this theorem, under Assumptions~\ref{assumption_1} and \ref{assumption_6}, the uplink SEs of UE~1 and UE~2 increase without bound as $M\to \infty$ even if the BS has only knowledge of the diagonal elements of the covariance matrices. A similar result can be proved for the downlink, using the methodology adopted in Appendix~D for proving Theorem~\ref{theorem:MMSE_precoding}. The details are omitted for space limitations.

\section{Asymptotic Spectral Efficiency in Multicell Massive MIMO}
\label{section:multi-user}

We will now generalize the results of Section~\ref{section:two-user} to a Massive MIMO network with $L$ cells, each comprising a BS with $M$ antennas and $K$ UEs.
There are $\taupu=K$ pilots and the $k$th UE in each cell uses the same pilot. Following the notation from \cite{hoydis2013massive}, the received signal ${\bf y}_j \in \mathbb{C}^{M}$ at BS $j$ is
\begin{equation}
{\bf y}_j = \sum_{l=1}^{L}  \sum_{i=1}^{K} \sqrt{\rho} \vect{h}_{jli} x_{li} + \vect{n}_j
\end{equation}
where $\rho$ is the normalized transmit power, $x_{li}$ is the unit-power signal from UE $i$ in cell $l$, $\vect{h}_{jli} \sim \CN (\vect{0}, \vect{R}_{jli})$ is the channel from this UE to BS $j$, $\vect{R}_{jli} \in \mathbb{C}^{M \times M}$ is the  channel covariance matrix, and $\vect{n}_j \sim \CN (\vect{0}, \vect{I}_{M})$ is the independent receiver noise at BS~$j$. Using a total uplink pilot power of $\rho^{\rm{tr}}$ per UE and standard MMSE estimation techniques \cite{hoydis2013massive}, BS $j$ obtains the estimate of $\vect{h}_{jli}$ as
\begin{align}
\hat{\vect{h}}_{jli} = \vect{R}_{jli} \vect{Q}_{ji}^{-1} \bigg( \sum_{l'=1}^{L} \vect{h}_{jl'i} + \frac{1}{\sqrt{\rho^{\rm{tr}}}} \vect{n}_{ji}   \bigg) \!\sim \!\CN \left( \vect{0},  \vect{\Phi}_{jli} \right)
\end{align}
where $\vect{n}_{ji} \sim \CN (\vect{0}, \vect{I}_{M})$ is noise, $\vect{Q}_{ji} = \sum_{l'=1}^{L} \vect{R}_{jl'i} + \frac{1}{\rho^{\rm{tr}}} \vect{I}_{M}$, and $
\vect{\Phi}_{jli}  = \vect{R}_{jli} \vect{Q}_{ji}^{-1} \vect{R}_{jli}$.
The estimation error $\tilde{\vect{h}}_{jli} = \vect{h}_{jli} - \hat{\vect{h}}_{jli}  \sim \CN \left( \vect{0}, \vect{R}_{jli}- \vect{\Phi}_{jli} \right)$ is independent of $\hat{\vect{h}}_{jli}$. However, the estimates $\hat{\vect{h}}_{j1i}, \ldots, \hat{\vect{h}}_{jLi}$ of the UEs with the same pilot are correlated as 
$
\mathbb{E}\{ \hat{\vect{h}}_{jni} \hat{\vect{h}}_{jmi}^{\Htran}\} = \vect{R}_{jni} \vect{Q}_{ji}^{-1} \vect{R}_{jmi}. 
$
\subsection{Uplink Data Transmission}
We denote by ${\bf v}_{jk} \in \mathbb {C}^{M}$ the receive combining vector associated with UE $k$ in cell $j$. Using the same technique as in \cite{hoydis2013massive,Marzetta2016a}, the uplink ergodic capacity is lower bounded by
\begin{equation} \label{eq:uplink-rate-expression-general}
\begin{split}
\mathsf{SE}_{jk}^{\rm {ul}} = \left( 1- \frac{\tau_p}{\tau_c} \right) \mathbb{E} \left\{ \log_2  \left( 1 + \gamma_{jk}^{\rm {ul}}  \right) \right\} \quad \textrm{[bit/s/Hz] }
\end{split}
\end{equation}
with the instantaneous effective SINR
\begin{align} \notag
\gamma_{jk}^{\rm {ul}} & =  \frac{ |  \vect{v}_{jk}^{\Htran} \hat{\vect{h}}_{jjk} |^2  }{{\mathbb{E}}\left\{ 
\!\sum\limits_{(l,i)\ne (j,k)} | \vect{v}_{jk}^{\Htran} {\vect{h}}_{jli} |^2
+| \vect{v}_{jk}^{\Htran} \tilde{\vect{h}}_{jjk} |^2+  \frac{\vect{v}_{jk}^{\Htran} \vect{v}_{jk}  }{\rho^{\rm {ul}}} 
\Big| \hat{\bf{h}}_{(j)}  \right\}} \\&= \frac{ |  \vect{v}_{jk}^{\Htran} \hat{\vect{h}}_{jjk} |^2  }{ 
 \vect{v}_{jk}^{\Htran}  \left(   \sum\limits_{(l,i)\ne (j,k)}  \hat{\vect{h}}_{jli} \hat{\vect{h}}_{jli}^{\Htran} +   \vect{Z}_j\right) \vect{v}_{jk}  
}   \label{eq:uplink-instant-SINR}
\end{align}
where ${\mathbb{E}}\{\cdot|{{\hat{\bf h}}_{(j)}}\}$ denotes the conditional expectation given the {MMSE} channel estimates available at BS $j$ and $\vect{Z}_j = \sum\nolimits_{l=1}^{L} \sum\nolimits_{i=1}^{K} (\vect{R}_{jli} - \vect{\Phi}_{jli}) +  \frac{1}{\rho^{\rm {ul}}}  \vect{I}_{M}$.
As shown in \cite{Ngo2012b,EmilEURASIP17}, the instantaneous effective {SINR} in \eqref{eq:uplink-instant-SINR} for {UE} $k$ in cell $j$ is maximized by 
\begin{equation} \label{eq:MMSE-combining}
\vect{v}_{jk} =  \Bigg(  \sum\limits_{l=1}^L\sum\limits_{i=1}^K \hat{\vect{h}}_{jli} \hat{\vect{h}}_{jli}^{\Htran} + \vect{Z}_j  \Bigg)^{\!-1}  \!\!  \hat{\vect{h}}_{jjk}.
\end{equation}
We refer to this ``optimal'' receive combining scheme as multicell MMSE (M-MMSE) combining. The ``multicell'' notion is used to differentiate it from the single-cell MMSE (S-MMSE) combining scheme \cite{hoydis2013massive,Guo2014a,Krishnan2014a}, which is widely used in the literature and defined as
\begin{align} 
\bar{\vect{v}}_{jk} = 
\left( \sum\limits_{i=1}^{K}  \hat{\vect{h}}_{jji} \hat{\vect{h}}_{jji}^{\Htran} + \bar{\vect{Z}}_j\right)^{-1}  \!\!  \hat{\vect{h}}_{jjk} \label{eq:S-MMSE-combining}
\end{align}
with $\bar{\vect{Z}}_j = \sum\nolimits_{i=1}^{K}\vect{R}_{jji} \!-\! \vect{\Phi}_{jji} + \sum\nolimits_{l=1,l \neq j}^{L}  \sum\nolimits_{i=1}^{K} \vect{R}_{jli} 
+  \frac{1}{\rho^{\rm {ul}}}     \vect{I}_{M}$.
The main difference from \eqref{eq:MMSE-combining} is that only channel estimates in the own cell are computed in S-MMSE, while 
$\hat{\vect{h}}_{jli}  \hat{\vect{h}}_{jli}^{\Htran} - \vect{\Phi}_{jli} $ is replaced with its average (i.e., zero) for all~$l \neq j$.
The computational complexity of S-MMSE is thus slightly lower than with M-MMSE (see \cite{EmilEURASIP17} for a detailed discussion).
However, both schemes only utilizes channel estimates that can be computed locally at the BS and the pilot overhead is identical since the same pilots are used to estimate both intra-cell and inter-cell channels. The S-MMSE scheme coincides with {M-MMSE} when there is only one isolated cell, but it is generally different and does not suppress interference from interfering {UEs} in other cells. Plugging \eqref{eq:MMSE-combining} into \eqref{eq:uplink-instant-SINR} yields
\begin{align} \label{eq:gammajk_MMSE}
\gamma_{jk}^{\rm {ul}}  &=  \hat{\vect{h}}_{jjk}^{\Htran}  \Bigg(  \sum\limits_{(l,i)\ne (j,k)}\hat{\vect{h}}_{jli} \hat{\vect{h}}_{jli}^{\Htran} + \vect{Z}_j  \Bigg)^{\!-1}\hat{\vect{h}}_{jjk}.\end{align}
We want to analyze $\gamma_{jk}^{\rm {ul}}$ when $M\to \infty$. To this end, we make the following two assumptions.
\begin{assumption}\label{assumption_3}As $M\to \infty$ $\forall j,l,i$, $\liminf_M \;\frac{1}{{M}}\tr ( \vect{R}_{jli}) > 0 $ and $
	\limsup_M \;\| \vect{R}_{jli}\|_2 < \infty$.
\end{assumption}
{
\begin{assumption}\label{assumption_4} For any UE $k$ in cell $j$ with ${\boldsymbol\lambda}_{jk} = [\lambda_{j1k}, \ldots,\lambda_{jLk}]^{\Ttran} \in \mathbb{R}^{L}$ and $\lalt=1,\ldots,L$ 
\begin{align}\label{Condition2_Assumption4_new}
	\liminf_M \inf_{\{{\boldsymbol\lambda}_{jk}: \, \lambda_{j \lalt k}=1\}} \frac{1}{{M}}\left\| \sum\limits_{l=1}^L\lambda_{jlk} \vect{R}_{jlk} \right\|_F^2 > 0.
\end{align}\end{assumption} 
The following is the fourth main result of the paper:
\begin{theorem} \label{theorem:M-MMSE}
If M-MMSE combining is used, then under Assumptions~\ref{assumption_3} and~\ref{assumption_4} the SINR $\gamma_{jk}^{\rm {ul}} $ increases a.s.~unboundedly as $M\to \infty$. Hence, $\mathsf{SE}_{jk}^{\rm {ul}}$ increases unboundedly as $M\to \infty$.
\end{theorem}
\begin{IEEEproof}
The proof is given in Appendix~F.
\end{IEEEproof}
This theorem proves the remarkable result that, under Assumptions \ref{assumption_3} and \ref{assumption_4}, the uplink SE of a multicell Massive MIMO network increases without bound as $M\to \infty$, despite pilot contamination. This is in sharp contrast to the finite limit in case of MR combining \cite{marzetta2010noncooperative} or any other single-cell combining scheme \cite{hoydis2013massive,Guo2014a,Krishnan2014a} and it is due to the fact that M-MMSE rejects the coherent interference caused by pilot contamination when Assumptions \ref{assumption_3} and \ref{assumption_4} hold. 
Note that these are the natural multicell generalizations of Assumptions~\ref{assumption_1} and~\ref{assumption_2}, respectively.
In particular, the condition \eqref{Condition2_Assumption4_new} says that the covariance matrices $\{\vect{R}_{jlk}: l=1,\ldots,L\}$ of the channels from  the pilot-sharing UEs to BS $j$ are asymptotically linearly independent, which implies the same condition for the estimated channels $\{\hat {\vect{h}}_{jlk}: l=1,\ldots,L\}$. This condition is used in Appendix~F to prove Theorem \ref{theorem:M-MMSE} in a fairly simple way.} However, we stress that Theorem \ref{theorem:M-MMSE} is valid also in a more general setting in which $\hat {\vect{h}}_{jjk}$ is asymptotically linearly independent of the estimates of all pilot-interfering UEs' channels, but some of the interfering channel estimates can be written as linear combinations of other interfering channels. Let $\mathcal S_{jk} \subseteq \{\hat {\vect{h}}_{jlk}: \forall l\ne j\}$ denote a subset of the estimated interfering channels that form a basis for all interfering channels. Under these circumstances, we only need to take the estimates in $\mathcal S_{jk}$ into account in the computation of the combining vector ${\bf v}_{jk}$ in \eqref{eq:MMSE-combining} and the same result follows. To gain further insights into this, we notice (as done for the two-user case in Section \ref{subsec:interpretation}) that one can find a receive combining vector that is orthogonal to the subspace spanned by $\mathcal S_{jk}$. This scheme exhibits an unbounded SE when $M\to \infty$ as it rejects the interference from all pilot-contaminating UEs (not only from those in $\mathcal S_{jk}$), while retaining an array gain that grows with $M$. We call this scheme multicell ZF (M-ZF) and define it as $\vect{v}_{jk} =  \hat{\vect{H}}_{jk} \big( \hat{\vect{H}}_{jk}^{\Htran}  \hat{\vect{H}}_{jk} \big)^{-1} \vect{e}_{1}$,
where $\vect{e}_{1}$ is the first column of $\vect{I}_{|\mathcal S_{jk}|+1}$ (with $|\mathcal S_{jk}|$ being the cardinality of $\mathcal S_{jk}$) and $\hat{\vect{H}}_{jk}\in \mathbb{C}^{N\times (|\mathcal S_{jk}|+1) }$
is the matrix with $\hat {\vect{h}}_{jjk}$ in the first column and the channel estimates in $\mathcal S_{jk}$ in the remaining columns. Since M-MMSE combining is the optimal scheme, it has to exhibit an unbounded SE if this is the case with M-ZF.

\subsection{Downlink Data Transmission}
During downlink data transmission, the BS in cell $l$ transmits $\vect{x}_l = \sqrt{\rho^{\rm{dl}}}\sum_{l=1}^{K} \vect{w}_{li} \varsigma_{li} $,
where  $\varsigma_{li} \sim \CN(0,1)$ is the data signal intended for UE~$i$ in the cell and $\rho^{\rm{dl}}$ is the normalized transmit power. This signal is assigned to a transmit precoding vector $ \vect{w}_{li} \in \mathbb{C}^{M}$, which satisfies $\mathbb{E} \{ \|  \vect{w}_{li} \|^2 \} =1$, such that 
$\mathbb{E} \{ \|  \vect{w}_{li} \varsigma_{li} \|^2 \} = \rho^{\rm{dl}}$ is the transmit power allocated to this UE. Using the same technique as in \cite{hoydis2013massive,Marzetta2016a},
the downlink ergodic channel capacity of UE $k$ in cell $j$ can be lower bounded by
$\mathsf{SE}_{jk}^{\rm {dl}} = \big( 1- \frac{\tau_p}{\tau_c} \big) \log_2   (  1 +
\gamma^{\mathrm{dl}}_{jk} )$ [bit/s/Hz] with 
\begin{equation} \label{eq:downlink-SINR-expression-forgetbound}
\gamma^{\mathrm{dl}}_{jk}=  \frac{ | \mathbb{E}\{  \vect{h}_{jjk}^{\Htran} \vect{w}_{jk}\} |^2  }{ 
\sum\limits_{l=1}^{L} \sum\limits_{i=1}^{K}  \mathbb{E} \{  | \vect{h}_{ljk}^{\Htran} \vect{w}_{li}  |^2 \}
-  | \mathbb{E}\{ \vect{h}_{jjk}^{\Htran}\vect{w}_{jk} \} |^2 + \frac{1}{\rho^{\rm{dl}}}  }. 
\end{equation}
Unlike $\gamma^{\mathrm{ul}}_{jk}$ in \eqref{eq:uplink-instant-SINR}, which only depends on the own combining vector ${\bf v}_{jk}$, $\gamma^{\mathrm{dl}}_{jk}$ depends on all precoding vectors $\{\vect{w}_{li}\}$. The precoding should ideally be selected jointly across the cells, which makes precoding optimization difficult in practice. Motivated by the uplink-downlink duality \cite{EmilEURASIP17}, it is reasonable to select $\{\vect{w}_{li}\}$ based on the M-MMSE combining vectors $\{\vect{v}_{jk}\}$ given by \eqref{eq:MMSE-combining}. This leads to M-MMSE precoding 
\begin{align}\label{eq:MMMSE_precoding}
{\bf w}_{jk}  = \sqrt{\vartheta_{jk}} \vect{v}_{jk} =  \sqrt{\vartheta_{jk}} \Bigg(  \sum\limits_{l=1}^L\sum\limits_{i=1}^K \hat{\vect{h}}_{jli} \hat{\vect{h}}_{jli}^{\Htran} + \vect{Z}_j  \Bigg)^{\!-1}    \hat{\vect{h}}_{jjk}
\end{align}
with the normalization factor $\vartheta_{jk} = (\sqrt{{\mathbb{E}}\left\{\|\vect{v}_{jk}\|^2\right\}})^{-1}$. 
This is the fifth main result of the paper:

\begin{theorem} \label{theorem:M-MMSE_precoding}
If M-MMSE precoding is used, then under Assumptions~\ref{assumption_3} and~\ref{assumption_4} the SINR $\gamma_{jk}^{\rm {dl}} $ grows unboundedly as $M\to \infty$. Hence, $\mathsf{SE}_{jk}^{\rm {dl}}$ grows unboundedly as $M\to \infty$.
\end{theorem}
\begin{IEEEproof}
Despite being much more involved, the proof basically unfolds from the same arguments used for proving Theorem~\ref{theorem:MMSE_precoding} and by exploiting the results of Appendix~F for Theorem~\ref{theorem:M-MMSE}.
\end{IEEEproof}

This theorem shows that an asymptotically unbounded downlink SE is achieved by all UEs in the network, despite the suboptimal assumptions of M-MMSE precoding, equal power allocation, and no estimation of the instantaneous realization of the precoded channels. The only important requirement is that the channel estimates to the desired UEs are asymptotically linearly independent from the channel estimates of pilot-contaminating UEs in other cells. Section~\ref{sec:numerical-results} demonstrates numerically that the DL SE grows without bound as $M \to \infty$.

\subsection{Approximate M-MMSE Combining and Precoding}\label{sec:approximate_MMSE}

In Section~\ref{sec:approximate_MMSE_two_user}, we have shown that the SE with the approximate M-MMSE scheme (that only utilizes the diagonals of the covariance matrices) grows unbounded as $M\to\infty$, in a two-user scenario. This result can be generalized to a multicell Massive MIMO network. Due to space limitations, we concentrate on the uplink. In particular, we assume that the signal of UE~$k$ in cell~$j$ is detected by using the approximate M-MMSE combining vector
\begin{equation} \label{eq:EW_MMSE-combining}
\vect{v}_{jk} =  \Bigg(  \sum\limits_{l=1}^L\sum\limits_{i=1}^K \hat{\vect{h}}_{jli} \hat{\vect{h}}_{jli}^{\Htran} + \vect{S}_j  \Bigg)^{\!-1}    \hat{\vect{h}}_{jjk}
\end{equation} 
where $\vect{S}_j =  \sum_{l=1}^{L}\sum_{i=1}^{K} \Big(\vect{D}_{jli} - \vect{D}_{jli}\boldsymbol{\Lambda}_{ji}^{-1}\vect{D}_{jli}\Big) + \frac{1}{\rho^{\rm {ul}}}  \vect{I}_M$ is a diagonal matrix and the EW-MMSE estimate of $\vect{h}_{jli}$ is
\begin{align}
\hat{\vect{h}}_{jli}  =  \frac{1}{ \sqrt{\rho^{\rm{tr}}} } {\bf D}_{jli} \boldsymbol{\Lambda}_{ji}^{-1} \left( \sum_{l'=1}^{L} \vect{h}_{jl'i} + \frac{1}{\sqrt{\rho^{\rm{tr}}}} \vect{n}_{ji}   \right) 
\end{align}
where $\vect{n}_{ji} \sim \CN (\vect{0}, \vect{I}_{M})$ is noise and ${\bf D}_{jli} \in \mathbb{R}^{M\times M}$ and $\boldsymbol{\Lambda}_{ji}\in \mathbb{R}^{M\times M}$ are diagonal with elements $\{[ \vect{R}_{jli} ]_{nn}: n=1,\ldots,M\}$ and $\{ \sum_{l'=1}^{L} [\vect{R}_{jl'i} ]_{nn} +  \frac{1}{ {\rho^{\rm{tr}}} }: n=1,\ldots,M\}$, respectively. 
Since ${\bf D}_{jli}$ and $\boldsymbol{\Lambda}_{ji}$ are diagonal, the computational complexity of EW-MMSE estimation is substantially lower than for MMSE estimation; see \cite{Shariati2014a} for details. Notice that the combining scheme in \eqref{eq:EW_MMSE-combining} can be applied without knowing the full channel covariance matrices, as it depends only on the diagonal elements of $\{{\bf R}_{jli}:l=1,\ldots,L\}$. This is because the elements of $\hat{\vect{h}}_{jli}$ are estimated separately, without exploiting the spatial channel correlation. By using the use-and-then-forget SE bound \cite{Marzetta2016a}, the uplink ergodic capacity of UE $k$ in cell $j$ can be lower bounded by ${\underline{\mathsf{SE}}}_{jk}^{\rm ul} = ( 1 - \frac{\taupu}{\tau_c}  )  \log_2  ( 1 + {\underline{\gamma}}_{jk}^{\rm ul}  )$ [bit/s/Hz] 
with 
\begin{align} \notag
&{\underline{\gamma}}_{jk}^{\rm ul}= \\& \frac{ | \mathbb{E}\{  \vect{v}_{jk}^{\Htran} \vect{h}_{jjk}\} |^2  }{ 
\sum\limits_{l=1}^{L} \sum\limits_{i=1}^{K}  \mathbb{E} \{  | \vect{v}_{jk}^{\Htran} \vect{h}_{jli}  |^2 \}
-  | \mathbb{E}\{ \vect{v}_{jk}^{\Htran}\vect{h}_{jjk} \} |^2 + \frac{1}{\rho^{\rm{ul}}}   \mathbb{E}\{ \left\|\vect{v}_{jk}\right\|^2 \}  }. 
\end{align}
We now want to understand how ${\underline{\gamma}}_{jk}^{\rm ul}$ behaves when $M\to \infty$ under the following assumption, which is the extension of Assumption~\ref{assumption_4} to the case where only the diagonals of covariance matrices are used for channel estimation and receive combining:
\begin{assumption}\label{assumption_7} For any UE $k$ in cell $j$ with ${\boldsymbol\lambda}_{jk} = [\lambda_{j1k}, \ldots,\lambda_{jLk}]^{\Ttran} \in \mathbb{R}^{L}$ and $\lalt=1,\ldots,L$ 
\begin{align}\label{Condition2_Assumption4}
	\liminf_M \inf_{\{{\boldsymbol\lambda}_{jk}: \, \lambda_{j \lalt k}=1\}} \frac{1}{{M}}\left\| \sum\limits_{l=1}^L\lambda_{jlk} \vect{D}_{jlk} \right\|_F^2 > 0.
\end{align}\end{assumption}
The following is the last main result of the paper:
\begin{theorem} \label{theorem:approximate_M-MMSE}
If approximate M-MMSE combining is used, then under Assumptions~\ref{assumption_3} and~\ref{assumption_7} the SINR ${\underline{\gamma}}_{jk}^{\rm ul}$ increases unboundedly as $M\to \infty$. Hence, $\underline{\mathsf{SE}}_{jk}^{\rm {ul}}$ increases unboundedly as $M\to \infty$.
\end{theorem}
\begin{IEEEproof}
The proof is omitted for space limitations, but follows along the lines of Theorem~\ref{theorem:EW-MMSE_precoding}.
\end{IEEEproof}

This theorem shows that it is sufficient that the diagonals of the covariance matrices are asymptotically linearly independent and known at the BS to achieve an unbounded uplink SE (and thus an unlimited capacity). This condition is generally satisfied since small random variations in the elements of the covariance matrices are sufficient to achieve asymptotic linear independence, as illustrated by Example~\ref{example2}. An unbounded SE can be also proved in the downlink  using similar methods (omitted for space reasons). This will be demonstrated numerically in the next section.

\section{Numerical Results}
\label{sec:numerical-results}

The simulation results can be reproduced using the code at \url{https://github.com/emilbjornson/unlimited-capacity}.
In this section, we will show numerically that an unlimited SE is achievable under pilot contamination. To this end, we first evaluate three ways to generate the channel covariance matrices and the resulting spatial correlation. For an arbitrary user, the covariance matrix $\vect{R}$ can be modeled by:

1) One-ring model for a ULA with half-wavelength antenna spacing and average large-scale fading $\beta$~\cite{Adhikary2013}. For an angle-of-arrival (AoA) $\theta$ and many scatterers that are uniformly distributed in the angular interval  $[\theta-\Delta,\theta+\Delta]$, the $(m,n)$th element of $\vect{R}$ is $[ \vect{R} ]_{m,n} = \frac{\beta}{2\Delta} \int_{-\Delta}^{\Delta} e^{ \pi \imath (n-m) \sin(\theta+\delta) }  d\delta$.

2)  Exponential correlation model for a ULA with correlation factor $r \in [0,1]$ between adjacent antennas, average large-scale fading $\beta$, and  AoA $\theta$ \cite{Loyka2001a}, which leads to $[ \vect{R} ]_{m,n} = \beta r^{|n-m|} e^{\imath  (n-m) \theta}$.

3)  Uncorrelated Rayleigh fading with average large-scale fading $\beta$ and independent log-normal large-scale fading variations over the array, which gives (similar to the perturbations considered in Example~\ref{example2}) 
  \begin{align} \label{eq:uncorrelated-fading-array-correlation-model}
\vect{R} = \beta \diag \left( 10^{f_1/10},\ldots, 10^{f_M/10}  \right)
\end{align}
where $f_m \sim \mathcal{N}(0,\sigma^2)$ and $\sigma$ denotes the standard deviation.

\begin{figure}[t!]
\begin{center}
\includegraphics[width=\columnwidth]{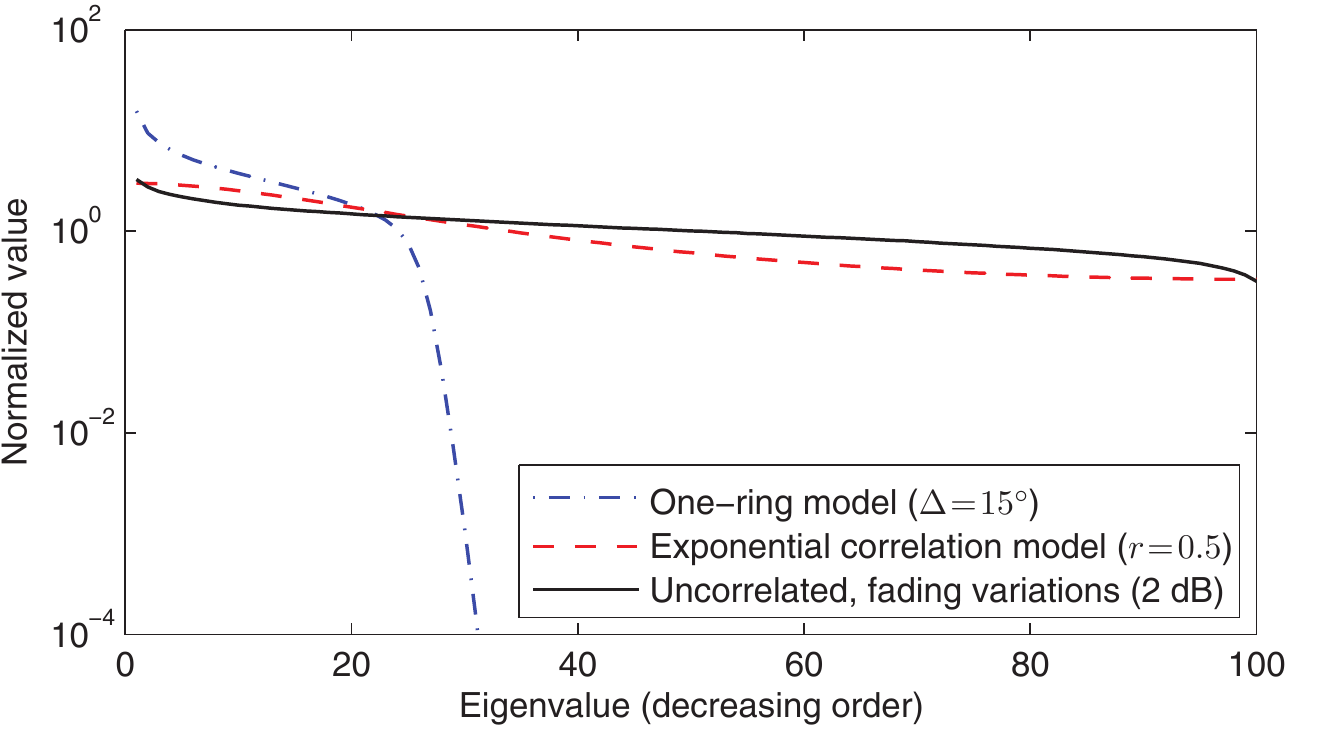}
\end{center}
\caption{Average eigenvalue distribution with $M=100$ and for three different channel covariance models, whereof one gives a rank-deficient covariance matrix and the others have full rank.} \label{figureEigenvalues} \vskip-4mm
\end{figure}

Fig.~\ref{figureEigenvalues} shows the eigenvalue distribution with the three covariance models above, 
for $M=100$ antennas, uniformly distributed AoAs $\theta$  in $[-\pi,+\pi)$, $\beta=1$, $\Delta =15^\circ$, $r=0.5$, and $\sigma=2$.
All three models create eigenvalue variations, but there are substantial differences. The one-ring model provides rank-deficient covariance matrices, where a large fraction of the eigenvalues is zero (this fraction is computed in~\cite{Adhikary2013}). In contrast, the other two models provide full-rank covariance matrices with more modest eigenvalue variations. In the remainder, we consider the latter two models to emphasize that our main results only require linear independence between the covariance matrices, not rank-deficiency (which in special cases give rise to orthogonal covariance supports~\cite{Yin2013a}).

\begin{figure}[t!]
\begin{center} 
\includegraphics[width=.75\columnwidth]{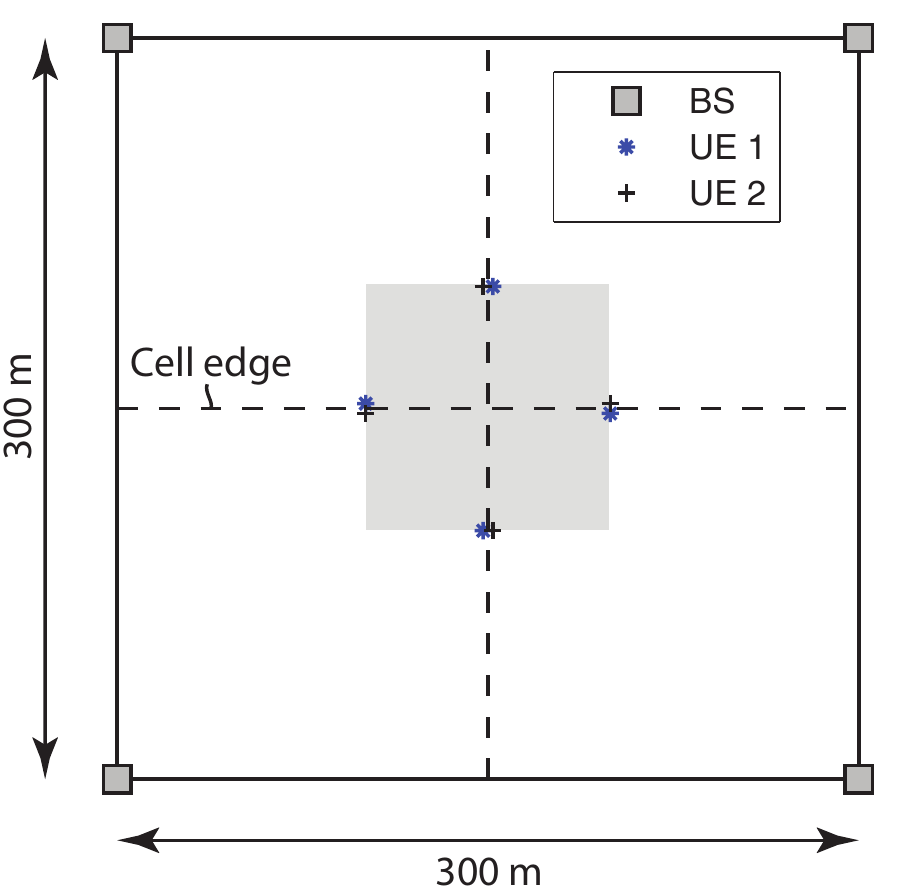}
\end{center} 
\caption{Multicell setup with two UEs per cell in the shaded cell-edge area. All UEs have similar AoAs to all BSs, which typically leads to similar covariance matrices and thus high pilot contamination.} \label{figureSetup}  
\end{figure}
 
\subsection{Uplink}

We consider the challenging symmetric setup in Fig.~\ref{figureSetup} with $L=4$ cells, $K=2$ UEs per cell, pilots of length $\taupu=K$, and coherence blocks of $\tau_c=200$ channel uses. The BSs are located at the four corners of the area and the UEs are all located at the cell edges and have similar but non-identical AoAs and distances to the BSs. Thus, the pilot contamination is very large in this setup. Note that the star-marked UEs share a pilot, while the plus-marked UEs share another pilot.

\begin{figure}[t!]
\begin{center}
\includegraphics[width=\columnwidth]{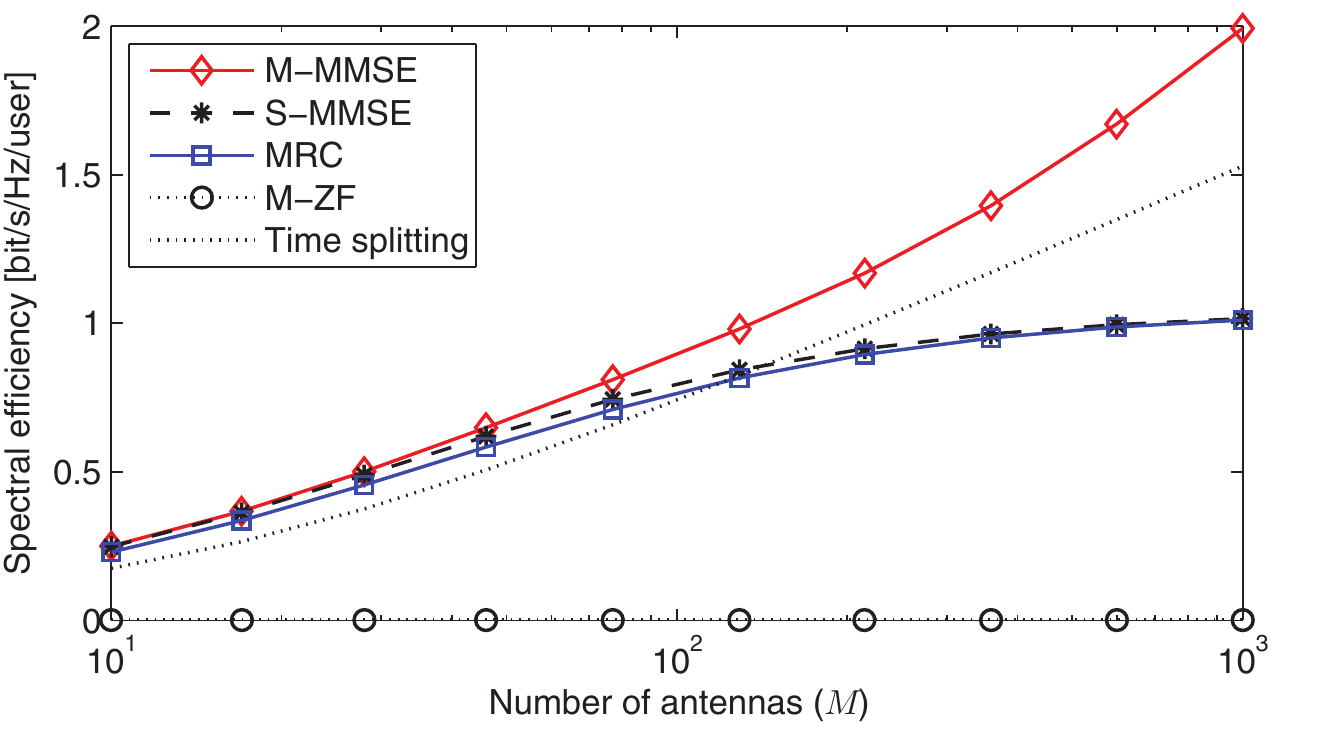}
\end{center}
\caption{Uplink SE as a function of $M$, for covariance matrices based on the exponential correlation model ($r=0.5$).} \label{figureAntennas} 
\end{figure}

The asymptotic behavior of the uplink SE is shown in Fig.~\ref{figureAntennas} using the exponential correlation model ($r=0.5$), with M-MMSE, S-MMSE, MR, and M-ZF, where the latter cancels interference between all UEs. The SE per UE is shown as a function of the number of antennas, in logarithmic scale.
The average SNR observed at a BS antenna is set equal in the pilot and data transmission: $\rho^{\rm{ul}} \tr(\vect{R}_{jli})/M =\rho^{\rm{tr}} \tr(\vect{R}_{jli})/M$. It is $-6.0$\,dB for the intracell UEs and between $-6.3$\,dB and $-11.5$\,dB for the interfering UEs in other cells.
Fig.~\ref{figureAntennas} shows that S-MMSE provides slightly higher SE than MR, but both converge to asymptotic limits of around 1\,bit/s/Hz as $M$ grows. In contrast, M-MMSE provides an SE that grows without bound.
The instantaneous effective SINR grows linearly with $M$, which is in line with Theorem~\ref{theorem:M-MMSE}, as seen from the fact that the SE grows linearly when the horizontal scale is logarithmic. M-ZF performs poorly because the channel estimates are so similar that full interference suppression removes most of the desired signal. In contrast, M-MMSE finds a non-trivial tradeoff between interference suppression and coherent combining of the desired signal, leading to superior SE. 
The reference curve ``time splitting'' considers the case when the 4 cells are active in different coherence blocks, to remove pilot contamination. MMSE combining is used and the SE grows without bound, but at a slower pace than with M-MMSE, due to the extra pre-log factor of $1/4$. Hence, even for a small system with $L=4$, it is inefficient to avoid pilot contamination by time splitting.

\begin{figure}
        \centering
        \begin{subfigure}[b]{\columnwidth} \centering 
                \includegraphics[width=\columnwidth]{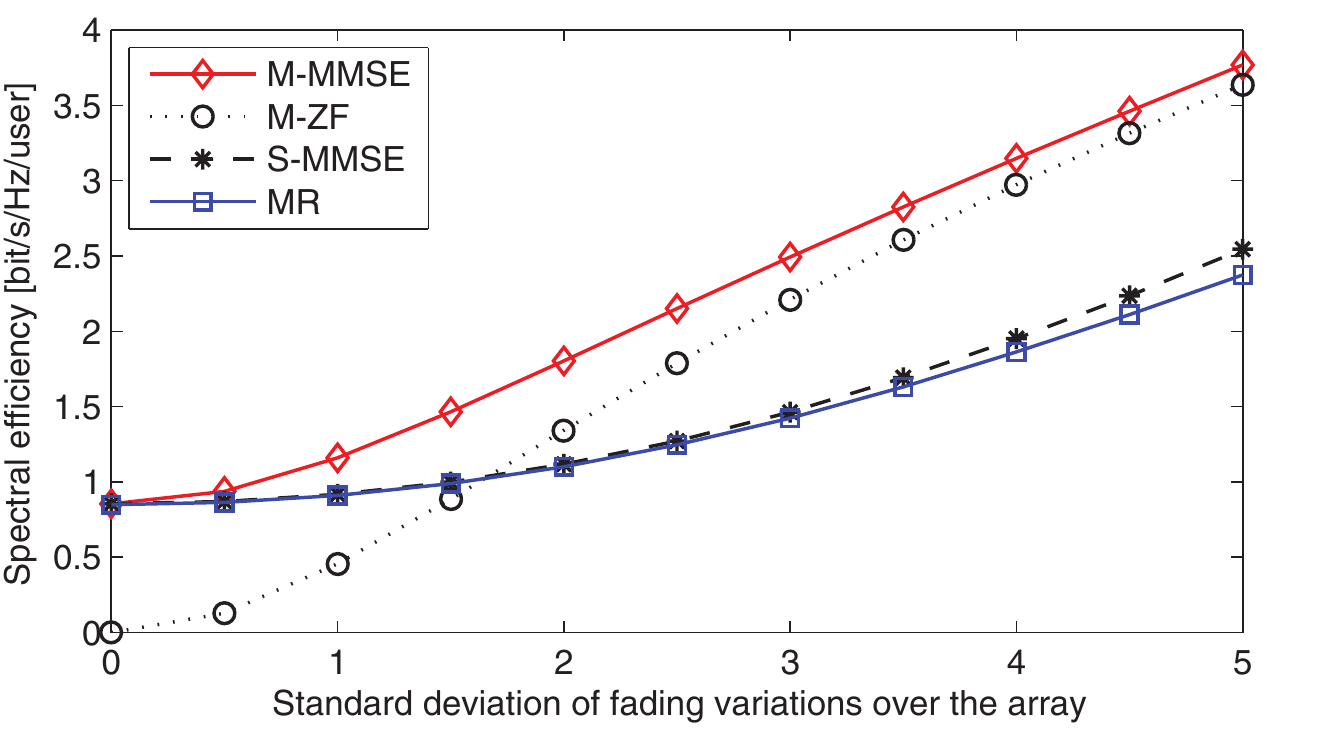} 
                \caption{Uplink SE}  
                \label{figureArrayFading:a}
        \end{subfigure}  
        \begin{subfigure}[b]{\columnwidth} \centering
                \includegraphics[width=\columnwidth]{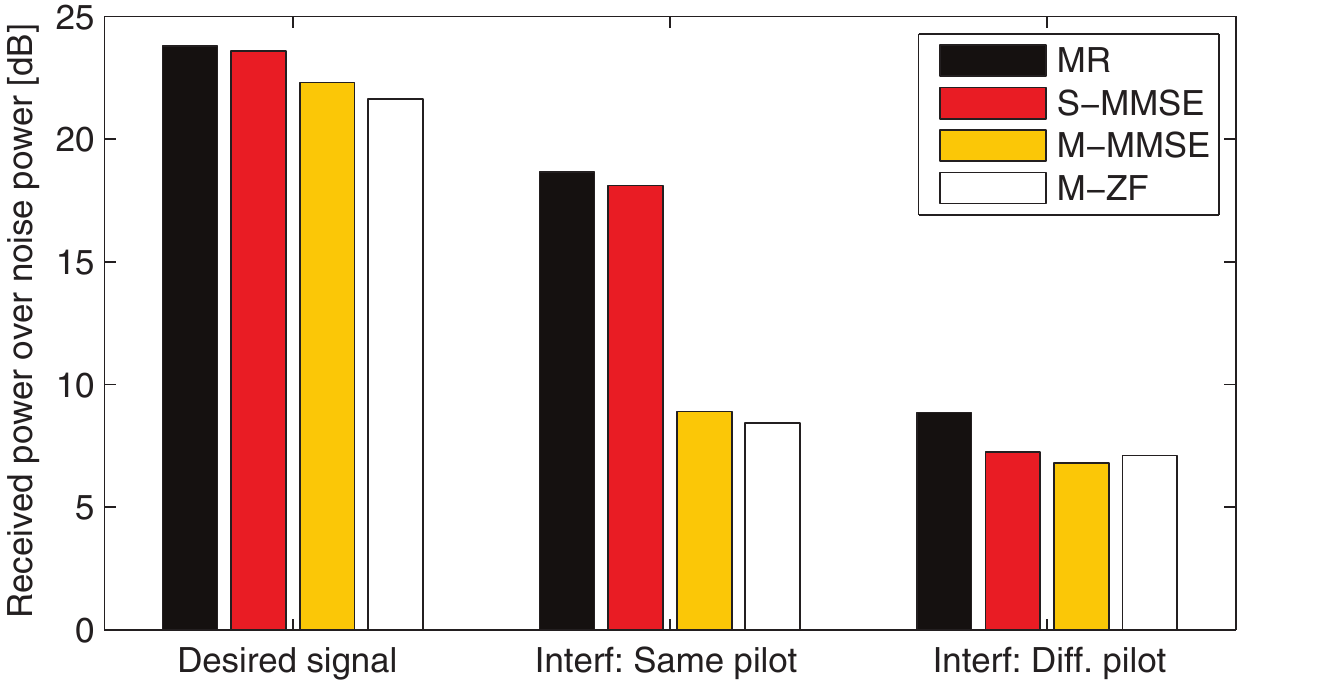} 
                \caption{Received signal power after receive combining} 
                \label{figureArrayFading:b} 
        \end{subfigure} 
        \caption{Uplink with covariance matrices modeled by \eqref{eq:uncorrelated-fading-array-correlation-model} for $M=200$ and $K=2$. (a) The SE as a function of the standard deviation $\sigma$ of the large-scale fading variations. (b) The received power after receive combining with $\sigma = 4$ is separated into desired signal power and interference from UEs with the same or different pilot than the desired UE.}
        \label{figureArrayFading}  
\end{figure}

Next, we consider the uncorrelated Rayleigh fading model in \eqref{eq:uncorrelated-fading-array-correlation-model} with independent large-scale fading variations over the array. The uplink SE with $M=200$ antennas and varying standard deviation $\sigma$ from $0$ to $5$ is shown in Fig.~\ref{figureArrayFading}(a). M-MMSE provides no benefit over S-MMSE or MR in the special case of $\sigma=0$, where the covariance matrices are linearly dependent (i.e., scaled identity matrices). This is a special case that has received massive attention in academic literature, mainly because it simplifies the mathematical analysis. However, M-MMSE provides substantial performance gains over S-MMSE and MR as soon as we depart from the scaled-identity model by adding small variations in the large-scale fading over the array, which make the covariance matrices linearly independent. This is in line with what we demonstrated in Example~\ref{example2}. As the variations increase, the SE with M-ZF improves particularly fast and approaches the SE with M-MMSE. M-ZF will never be the better scheme since M-MMSE is optimal. The motivation behind this simulation is the measurement results reported in \cite{Gao2015b}, which show large-scale variations of around 4\,dB over a massive MIMO array---this corresponds to $\sigma \approx 4$ in our setup.

Fig.~\ref{figureArrayFading}(b) shows the received power (normalized by the noise power) after receive combining for an arbitrary UE when $\sigma = 4$. It is divided into the desired signal power, the interference from UEs using the same pilot, and the interference from UEs using a different pilot. The figure shows that MR and S-MMSE suffer from strong interference from the UEs that use the same pilot, since these schemes are unable to mitigate the coherent interference caused by pilot contamination. In contrast, M-MMSE and M-ZF mitigate all types of interference and receive roughly the same amount of interference from UEs with the same or different pilots. Note that the price to pay for the interference rejection is a reduction in desired signal power when using M-MMSE and M-ZF.
 
\subsection{Downlink}

The setup in Fig.~\ref{figureSetup} is also used in the downlink wherein we set $\rho^{\rm{dl}}=\rho^{\rm{ul}}$ to get the same SNRs as in the uplink.
We consider a setup with both spatial channel correlation and large-scale fading variations over the array, such that the EW-MMSE estimator is suboptimal but Assumption~\ref{assumption_7} is satisfied.
More precisely, we consider a combination of the exponential correlation model  and \eqref{eq:uncorrelated-fading-array-correlation-model}: $[ \vect{R} ]_{m,n} = \beta r^{|n-m|} e^{\imath  (n-m) \theta} 10^{(f_m +f_n) /20}$, where $\theta$ is the AoA, $r=0.5$ is used as correlation factor, and $f_1,\ldots,f_M \sim \mathcal{N}(0,\sigma^2)$ give independent large-scale fading variations over the array with $\sigma=4$.

\begin{figure} 
        \centering
        \begin{subfigure}[b]{\columnwidth} \centering 
                \includegraphics[width=\columnwidth]{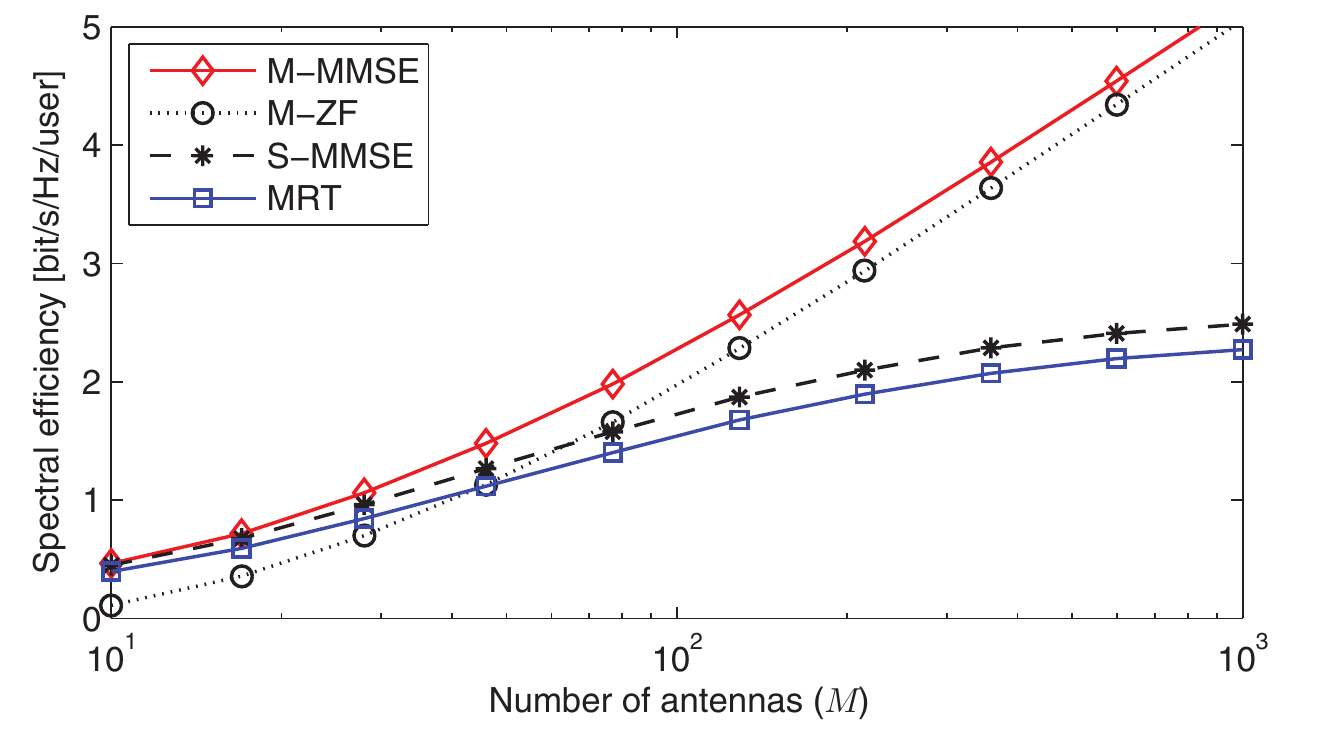} 
                \caption{MMSE estimation} 
                \label{figureDownlinkSimulation:a}
        \end{subfigure} 
        \begin{subfigure}[b]{\columnwidth} \centering
                \includegraphics[width=\columnwidth]{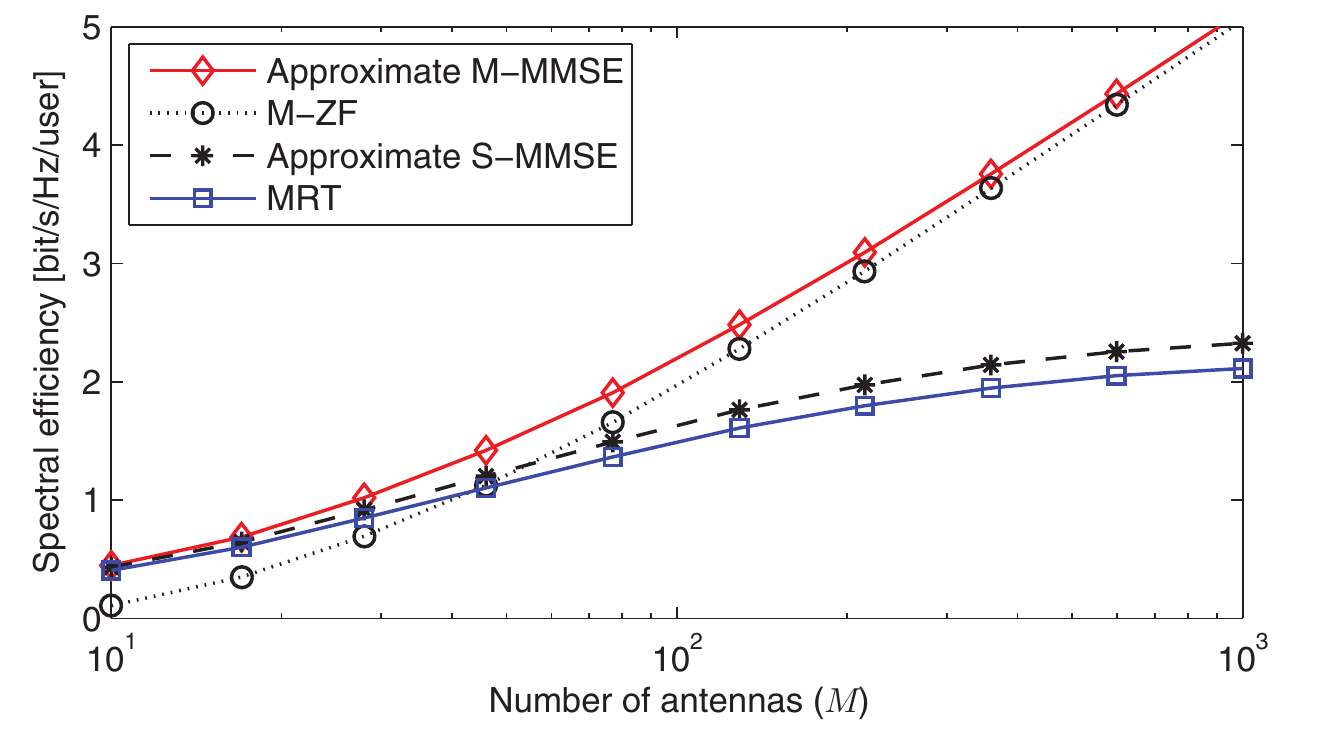} 
                \caption{EW-MMSE estimation}  
                \label{figureDownlinkSimulation:b} 
        \end{subfigure} 
        \caption{Downlink SE as a function of $M$ for $K=2$, when using either the MMSE estimator (with full covariance knowledge) or the EW-MMSE estimator (with known diagonals of the covariance matrices). The exponential correlation model with $r=0.5$ is used, but with large-scale fading variations over the array with $\sigma=4$.}
        \label{figureDownlinkSimulation}  
\end{figure}

The downlink SE is shown in Fig.~\ref{figureDownlinkSimulation} as a function of $M$, where Fig.~\ref{figureDownlinkSimulation}(a) shows results with the MMSE estimator that uses the full channel covariance matrices and Fig.~\ref{figureDownlinkSimulation}(b) shows results with the EW-MMSE estimator that only uses the diagonals of the covariance matrices. When using the EW-MMSE estimator, we consider the approximate M-MMSE scheme in \eqref{eq:EW_MMSE-combining} and a corresponding approximation of S-MMSE, while M-ZF and MR are as before.
The results in Fig.~\ref{figureDownlinkSimulation}(a) with the MMSE estimator are similar to the uplink in Fig.~\ref{figureArrayFading}(a): M-MMSE and M-ZF provide SEs that grow without bound, while the SEs with S-MMSE and MR converge to finite limits. In contrast to the uplink, M-MMSE and M-ZF precoding are both suboptimal in the downlink, but they can be shown to be asymptotically equal.\footnote{For M-MMSE precoding in \eqref{eq:MMMSE_precoding}, $\vect{Z}_j$ has bounded spectral norm while $\sum_l \sum_i \hat{\vect{h}}_{jli} \hat{\vect{h}}_{jli}^{\Htran}$ has $LK$ eigenvalues that grow unboundedly as $M \to \infty$. As the impact of $\vect{Z}_j$ vanishes, the approach in \cite{Bjornson2014d} can be used to prove that M-MMSE approaches M-ZF asymptotically.} Interestingly, the same behaviors are observed in Fig.~\ref{figureDownlinkSimulation}(b) when using the EW-MMSE estimator, which is a suboptimal estimator that neglects the off-diagonal elements of the covariance matrices.
This result is in line with Theorem~\ref{theorem:approximate_M-MMSE}. There is a small SE loss (2\%--4\% for M-MMSE) compared to Fig.~\ref{figureDownlinkSimulation}(a), but this is a minor price to pay for the greatly simplified acquisition of covariance information (estimating the entire diagonal is as simple as estimating a single parameter \cite{Shariati2014a,Bjornson2016c}).

\begin{figure} 
        \centering
        \begin{subfigure}[b]{\columnwidth} \centering 
                \includegraphics[width=\columnwidth]{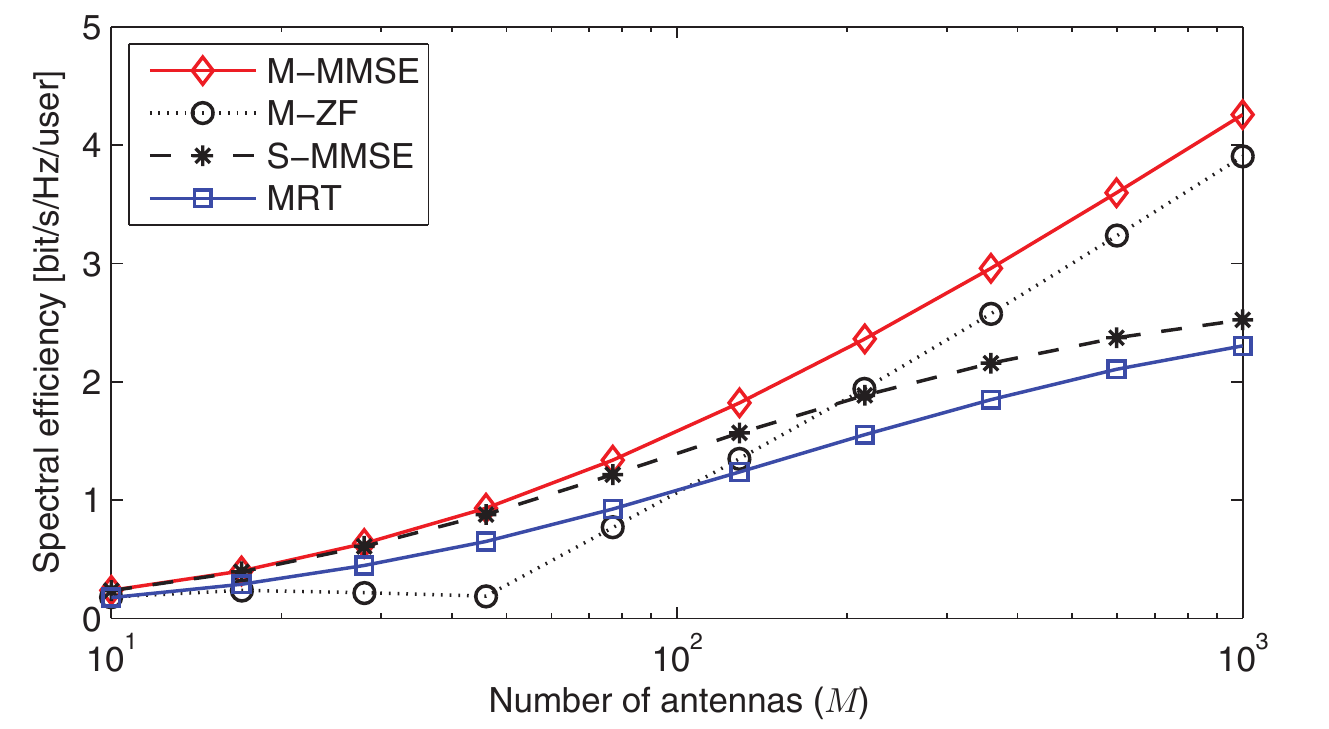} 
                \caption{MMSE estimation} 
                \label{figureDownlinkSimulation:a}
        \end{subfigure}  \hspace{3mm}
        \begin{subfigure}[b]{\columnwidth} \centering
                \includegraphics[width=\columnwidth]{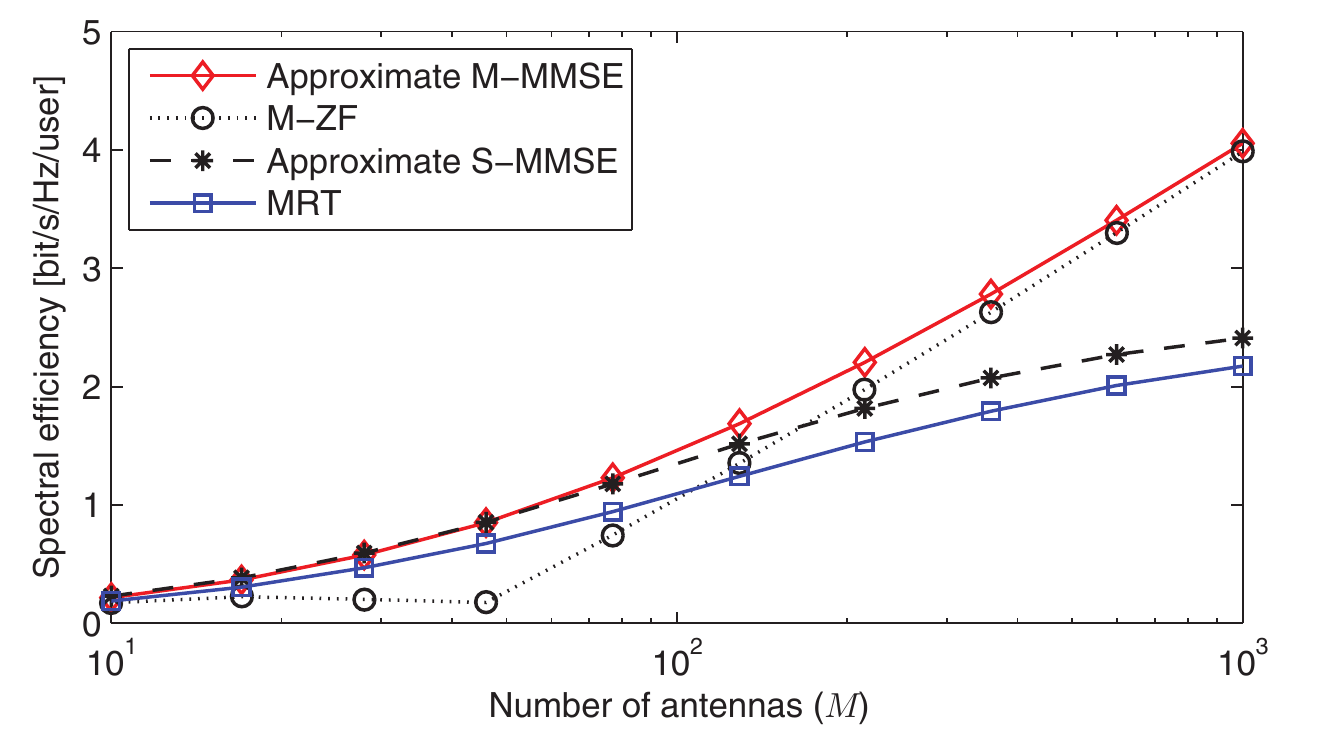} 
                \caption{EW-MMSE estimation}  
                \label{figureDownlinkSimulation:b} 
        \end{subfigure} 
        \caption{Downlink SE as a function of $M$ for $K=10$ UEs that are uniformly distributed in the shaded cell edge area. The setup and covariance model are otherwise the same as in Fig.~\ref{figureDownlinkSimulation}.}
        \label{figureDownlinkSimulationK10} 
\end{figure}

We now increase the number of UEs per cell to $K=10$, which leads to more interference but the same pilot contamination per UE. The UEs are uniformly and independently distributed in the cell-edge area, which is the shaded area in Fig.~\ref{figureSetup}. The channel model is the same as in the previous figure. The downlink SE per UE is shown in Fig.~\ref{figureDownlinkSimulationK10} when using either MMSE or EW-MMSE estimation. The results resemble the ones for $K=2$, but the curves are basically shifted to the right due to the additional interference. M-MMSE and M-ZF provide SEs that grow without bound, while the SE with S-MMSE and MR saturate, but more antennas are needed before reaching saturation.

\section{Conclusions and Practical Implications}
\label{section:conclusion}

We proved that the capacity of Massive MIMO systems increases without bound as $M \to \infty$ in the presence of pilot contamination, despite the previous results that pointed toward the existence of a finite limit. This was achieved by showing that the conventional lower bounds on the capacity increase without bound when using M-MMSE precoding/combining. These schemes exploit the fact that the MMSE channel estimates of UEs that use the same pilot are linearly independent, due to their generally linearly independent covariance matrices. For our results to hold, the covariance matrices can have full rank and minor eigenvalue variations are sufficient. 
There are special cases where the channel covariance matrices are linearly dependent, but these are not robust to minor perturbations of the covariance matrices. Hence, they are anomalies that will never appear in practice or be drawn from a random distribution, although they have frequently been studied in the academic literature. Since the SE of MR (also known as conjugate beamforming or matched filtering) generally has a finite limit, we conclude that this scheme is not asymptotically optimal in Massive MIMO. Note that our results do not imply that the pilot contamination effect disappears; there is still a performance loss caused by estimation errors and interference rejection, but there is no fundamental capacity limit.

Most of our results assume that the full covariance matrices of the channels are known, but this is not a critical requirement. 
Theorems~\ref{theorem:EW-MMSE_precoding} and \ref{theorem:approximate_M-MMSE} proved that it is sufficient that the diagonals of the covariance matrices are known and linearly independent between pilot-sharing UEs; a condition that has been shown to hold for practical channels by the measurements in \cite{Gao2015b}. Such statistical information can be accurately estimated from only some tens of channel observations \cite{Bjornson2016c}, whereof some contain the desired signal plus interference/noise and some contain only interference/noise.

The purpose of analyzing the asymptotic capacity when $M \to \infty$ is not that we advocate the deployment of BSs with a nearly infinite number of antennas---that is physically impossible in a finite-sized world and the conventional channel models will eventually break down since more power is received than was transmitted.
The importance of asymptotics is instead what it tells us about practical networks with finite numbers of antennas.
For example, consider a network with any finite number of UEs that each have a finite-valued data rate requirement. Our main results imply that we can always satisfy these requirements by deploying sufficiently many antennas, even in the presence of pilot contamination. In fact, it is enough to have two channel uses per coherence block (one for pilot, one for data) to deliver any capacity value to any finite number of UEs. The linear M-MMSE scheme is sufficient to achieve this in practice and interference can be treated as noise in the receivers, because the capacity lower bounds that we considered rely on such simplifications.

\section*{Appendix A -- Useful Results} \label{appendix:useful-results}

\begin{lemma}[Theorem 3.4, Corollary~3.4 \cite{Couillet_book}] \label{lemma3}
Let ${\bf A} \in\mathbb{C}^{M\times M}$ and ${\bf x},{\bf y}\sim \CN ({\bf 0}, \frac{1}{M} {\bf I}_M)$. Assume that ${\bf A}$ has uniformly bounded spectral norm and that ${\bf x}$ and ${\bf y}$ are mutually independent and independent of ${\bf A}$. Then, $ {\bf x}^{\Htran}{\bf A}{\bf x} \asymp \frac{1}{M} \tr ({\bf A})$, ${\bf x}^{\Htran}{\bf A}{\bf y} \asymp 0$ and ${\mathbb{E}}\{|{\bf x}^{\Htran}{\bf A}{\bf x} - \frac{1}{M} \tr ({\bf A})|^p\}= \mathcal {O}({M^{-p/2}})$.
\end{lemma}
\begin{lemma} [\!\cite{Marshall2011}]\label{lemma2}
For any positive semi-definite $M \times M$ matrices $\vect{A}$ and $\vect{B}$, it holds that $\frac{1}{M}\tr \left( \vect{A} \vect{B}\right) \le \| \vect{A} \vect{B} \|_2\le\| \vect{A} \|_2\|\vect{B} \|_2$, $\tr \left( \vect{A} \vect{B}\right) \le \| \vect{A} \|_2\tr \left( \vect{B}\right)$ and $ \tr \left( (\vect{I}+\vect{A})^{-1} \vect{B} \right) \geq \frac{1}{1+ \| \vect{A}  \|_2}  \tr (\vect{B})$.
\end{lemma}

\begin{lemma} [Matrix inversion lemma]\label{MIL}
Let ${\bf A} \in\mathbb{C}^{M\times M}$ be a Hermitian invertible matrix, then for any vector ${\bf x}\in \mathbb{C}^{M}$ and any scalar $\rho \in \mathbb{C}$ such that ${\bf A} + \rho {\bf x}{\bf x}^{\Htran}$ is invertible ${\bf x}^{\Htran}({\bf A} + \rho {\bf x}{\bf x}^{\Htran})^{-1} = \frac{{\bf x}^{\Htran}{\bf A}^{-1}}{1 + \rho {\bf x}^{\Htran}{\bf A}^{-1}{\bf x}}$ and $({\bf A} + \rho {\bf x}{\bf x}^{\Htran})^{-1} = {\bf A}^{-1} - \frac{\rho{\bf A}^{-1}{\bf x}{\bf x}^{\Htran}{\bf A}^{-1}}{1 + \rho {\bf x}^{\Htran}{\bf A}^{-1}{\bf x}}$.

Let ${\bf U}, {\bf C},{\bf V}$ be matrices of compatible sizes, then if ${\bf C}$ is invertible $\left({\bf A} + {\bf U}{\bf C}{\bf V}\right)^{-1} = {\bf A}^{-1}-{\bf A}^{-1}{\bf U}\left({\bf C}^{-1} + {\bf V}{\bf A}^{-1}{\bf U}\right)^{-1}{\bf V}{\bf A}^{-1}$.
\end{lemma}

\section*{Appendix B -- Proof of Theorem~\ref{theorem:MMSE}} \label{appendix:proof-theorem:MMSE}

By applying Lemma~\ref{MIL}, we may rewrite $\gamma_1^{\rm{ul}}$ in \eqref{eq:gamma1_MMSE} as 
\begin{align}\label{eq:gamma_1.1}
\gamma_{1}^{\rm {ul}} = {M}\Bigg(\frac{1}{M}\hat{\vect{h}}_{1}^{\Htran}{\bf Z}^{-1}\hat{\vect{h}}_{1} - \frac{\Big|\frac{1}{M}\hat{\vect{h}}_{1}^{\Htran}{\bf Z}^{-1}\hat{\vect{h}}_{2}\Big|^2}{\frac{1}{M}+\frac{1}{M}\hat{\vect{h}}_{2}^{\Htran}{\bf Z}^{-1}\hat{\vect{h}}_{2}}\Bigg)
\end{align}
by also multiplying and dividing each term by $M$.
Under Assumption~\ref{assumption_1} and using Lemma~\ref{lemma3} we have, as ${M \to \infty}$, that\footnote{Under Assumption~\ref{assumption_1}, ${\bf Q}^{-1} {\bf R}_i{\bf Z}^{-1}{\bf R}_k$ has uniformly bounded spectral norm, which can be easily proved using Lemma~\ref{lemma2}.} 
\begin{align}\label{eq:beta_11}
\frac{1}{M}\hat{\bf h}_1^{\Htran}{\bf Z}^{-1}\hat{\bf h}_1 &\asymp \frac{1}{M}\tr ( \vect{\Phi}_{1}{\bf Z}^{-1} )\triangleq \beta_{11}\\\label{eq:beta_22}
\frac{1}{M}\hat{\bf h}_2^{\Htran}{\bf Z}^{-1}\hat{\bf h}_2 &\asymp\frac{1}{M}\tr ( \vect{\Phi}_{2}{\bf Z}^{-1} ) \triangleq \beta_{22}\\ \label{eq:beta_12}
\frac{1}{M}\hat{\bf h}_1^{\Htran}{\bf Z}^{-1}\hat{\bf h}_2 &\asymp\frac{1}{M}\tr (\vect{\Upsilon}_{12} {\bf Z}^{-1} ) \triangleq \beta_{12}.
\end{align}
Note that $\beta_{11}$, $\beta_{22}$, and $\beta_{12}$ are non-negative real-valued scalars, since the trace of a product of positive semi-definite matrices is always non-negative. Using this notation, it follows from Assumption~\ref{assumption_1} that\footnote{{This can be proved by similar arguments as in Appendix~C, since $\tr ({\bf A}^2)\ge (\tr ({\bf A}))^2/\rm{rank}({\bf A})$ if ${\bf A}$ is Hermitian and ${\bf A}\ne {\bf 0}$.}} $\liminf_M\beta_{22}>0$ and we obtain
\begin{align}
\frac{\gamma_{1}^{\rm {ul}}}{M} \asymp \delta_1 \triangleq \beta_{11} - \frac{\beta_{12}^2}{ \beta_{22}}.
\label{eq:asympotitic_SINR}
\end{align}
To proceed, notice that Assumption~\ref{assumption_2} implies the following result, as proved in Appendix C.

\begin{corollary} \label{cor:assumption3}
If Assumption~\ref{assumption_2} holds, then for $\boldsymbol{\lambda} = [\lambda_1, \lambda_2]^{\Ttran} \in \mathbb{R}^2$ and $i=1,2$,
\begin{align} \notag
&\mathop {\liminf}\limits_M \inf_{\{\boldsymbol{\lambda}: \, \lambda_i=1\}} \\ &\frac{1}{{M}}\tr \Big( \vect{Q}^{-1}\big(\lambda_1\vect{R}_{1} +\lambda_2 \vect{R}_{2}\big) \vect{Z}^{-1} \big(\lambda_1\vect{R}_{1} +\lambda_2 \vect{R}_{2}\big)  \Big) > 0. \label{eq:assumption_2}
\end{align}
\end{corollary}
By expanding the condition in Corollary \ref{cor:assumption3} for $i=1$, we have that
\begin{align}\label{eq:Appendix_B.1_3}
	\liminf_M \inf_{\lambda_2}\left(\beta_{11} + {\lambda_2^2}\beta_{22} + 2\lambda_2\beta_{12}\right) >0.
	\end{align}
By the definition of the $\liminf_M$ operator, $\liminf_M \beta_{22}>0$ 
holds if and only if every convergent subsequence has a non-zero limit, i.e., $\lim_M \beta_{22}  >0$. This ensures that, for an arbitrary convergent subsequence, 
\begin{align} \label{eq:Appendix_B.1_4}
\inf_{\lambda_2}\left(\beta_{11} + {\lambda_2^2}\beta_{22} + 2\lambda_2\beta_{12}\right) = \beta_{11} - \frac{\beta_{12}^2}{\beta_{22}} = \delta_1
	\end{align}
where the infimum is attained by $\lambda_2 = \beta_{12}/\beta_{22}$. Substituting \eqref{eq:Appendix_B.1_4} into \eqref{eq:Appendix_B.1_3}, implies that $\liminf_M\delta_1>0$. Therefore, we have that $\gamma_{1}^{\rm {ul}}$ grows a.s.~unboundedly and, thus, the first part of the theorem follows.

Since $\gamma_{1}^{\rm{ul}} $  grows a.s.~unboundedly and the logarithm is a strictly increasing function, it follows that $\log_2(1+\gamma_{1}^{\rm{ul}})$ also grows a.s.~without bound. Moreover, since the almost sure divergence of a  sequence of non-negative random variables implies the divergence of its expected value, it follows that also $\mathsf{SE}_{1}^{\rm {ul}} = (1-\taupu/\tau_c) \mathbb{E} \left\{ \log_2  \left( 1 + \gamma_{1}^{\rm {ul}}  \right) \right\}$ grows without bound.

\section*{Appendix C -- Proof of Corollary~\ref{cor:assumption3} in Appendix B} \label{appendix:proof:cor:assumption3}

Consider $i=1$ and notice that the argument on the left-hand side of \eqref{eq:assumption_2} is lower bounded as
\begin{align} 
\frac{ \frac{1}{M}\| \vect{R}_{1} + \lambda_2 \vect{R}_{2} \|_F^2 }{ ( \frac1{\rho^{\rm{tr}}} + \| \vect{R}_{1} + \vect{R}_{2}\|_2) ( \frac1{\rho^{\rm{ul}}} + \| \sum_{k=1}^{2} (\vect{R}_{k} - \vect{\Phi}_{k}) \|_2)   }   \label{eq:assumption_2_relaxed-derivation}
\end{align}
by applying Lemma~\ref{lemma2} twice. The denominator of \eqref{eq:assumption_2_relaxed-derivation} is bounded from above due to Assumption~\ref{assumption_1} and independent of $\lambda_2$. This proves that Assumption~\ref{assumption_2} is sufficient for \eqref{eq:assumption_2} to hold for $i=1$. The result for $i=2$ follows by interchanging the indices in the proof.

\section*{Appendix D -- Proof of Theorem~\ref{theorem:MMSE_precoding}} \label{appendix:proof-theorem:MMSE_precoding}

We begin by plugging \eqref{eq:Section3_precoding} into \eqref{eq:gamma1_DL} to obtain
\begin{align} \label{eq:gamma1_DL_11}
\gamma_{1}^{\rm {dl}}  &=  \frac{ |  {\mathbb{E}}\left\{\vect{h}_{1}^{\Htran} {\vect{v}}_{1}\right\} |^2  }{ \frac{\vartheta_2}{\vartheta_1}{\mathbb{E}}\left\{| \vect{h}_{1}^{\Htran} {\vect{v}}_{2}|^2\right\} + {\mathbb{V}} \{ {\vect{h}}_{1}^{\Htran} {\vect{v}}_{1} \} + \frac{1}{\rho^{\rm dl}\vartheta_1}}.
\end{align}
We need to characterize all the terms in \eqref{eq:gamma1_DL_11} and begin with ${\mathbb{E}}\left\{\vect{h}_{1}^{\Htran} {\vect{v}}_{1}\right\}$. Notice that ${\mathbb{E}}\left\{\vect{h}_{1}^{\Htran} {\vect{v}}_{1}\right\}={\mathbb{E}}\big\{\hat {\vect{h}}_{1}^{\Htran} {\vect{v}}_{1}\big\}$ since ${\vect{v}}_{1}$ is independent of the zero-mean error $\tilde{\vect{h}}_{1}$. Then, we can express $\hat{\vect{h}}_{1}^{\Htran} {\vect{v}}_{1}$ as 
\begin{align}\label{eq:C.1}
\hat{\vect{h}}_{1}^{\Htran} {\vect{v}}_{1}  =\frac{{\hat{\vect{h}}}_{1}^{\Htran}\left(\hat{\vect{h}}_{2} \hat{\vect{h}}_{2}^{\Htran} + \vect{Z}  \right)^{-1} \hat{\vect{h}}_{1}}{{1 + {\hat{\vect{h}}}_{1}^{\Htran}\left(\hat{\vect{h}}_{2} \hat{\vect{h}}_{2}^{\Htran} + \vect{Z}  \right)^{-1} \hat{\vect{h}}_{1}}}=\frac{\gamma_{1}^{\rm {ul}}}{1 + {\gamma_{1}^{\rm {ul}}}}\end{align}
by first applying Lemma \ref{MIL} and then identifying $\gamma_{1}^{\rm {ul}}$ in 
\eqref{eq:gamma1_MMSE} in the numerator and denominator. Theorem~\ref{theorem:MMSE} proves that $\frac{\gamma_{1}^{\rm {ul}}}{M} \asymp \delta_1$ and applying this result to \eqref{eq:C.1} yields $\hat{\vect{h}}_{1}^{\Htran} {\vect{v}}_{1}\asymp 1$. By the dominated convergence theorem and the continuous mapping theorem \cite{Couillet_book}, we then have that $|{\mathbb{E}}\{\vect{h}_{1}^{\Htran} {\vect{v}}_{1}\}|^2   \asymp1$.

Consider now the noise term $\frac{1}{\rho^{\rm dl}\vartheta_1} =\frac{{\mathbb{E}}\{\|\vect{v}_1\|^2\}}{\rho^{\rm dl}}$ where $\vartheta_1 = ({\mathbb{E}}\left\{\|\vect{v}_1\|^2\right\})^{-1}$. By applying Lemma \ref{MIL} twice, we may rewrite $\|\vect{v}_1\|^2$ as \begin{align}\notag
\|\vect{v}_1\|^2 &= \frac{\hat{\vect{h}}_{1}^{\Htran}\left(\hat{\vect{h}}_{2} \hat{\vect{h}}_{2}^{\Htran} + \vect{Z}  \right)^{-2} \hat{\vect{h}}_{1} }{\left(1 +\gamma_{1}^{\rm {ul}}\right)^2} \\&= \frac{1}{M} \frac{\frac{1}{M} \hat{\vect{h}}_{1}^{\Htran}\left(\hat{\vect{h}}_{2} \hat{\vect{h}}_{2}^{\Htran} + \vect{Z}  \right)^{-2} \hat{\vect{h}}_{1} }{\left(\frac{1}{M} + \frac{1}{M} \gamma_{1}^{\rm {ul}}\right)^2}.\label{eq:C_10}
\end{align}
Let ${\rm{Re}}(\cdot)$ denote the real-valued part of a scalar. The numerator in \eqref{eq:C_10} can be expressed as
\begin{align}\notag\frac{1}{M}\hat{\vect{h}}_{1}^{\Htran}{\bf Z}^{-2}\hat{\vect{h}}_{1} &- 2\frac{{\rm{Re}}(\frac{1}{M}\hat{\vect{h}}_{1}^{\Htran}{\bf Z}^{-1}\hat{\vect{h}}_{2}\frac{1}{M}\hat{\vect{h}}_{2}^{\Htran}{\bf Z}^{-2}\hat{\vect{h}}_{1})}{\frac{1}{M} +  \frac{1}{M}\hat{\bf h}_2^{\Htran}{\bf Z}^{-1}\hat{\vect{h}}_{2}} \\&+ \frac{\frac{1}{M}\hat{\vect{h}}_{2}^{\Htran}{\bf Z}^{-2}\hat{\vect{h}}_{2}|\frac{1}{M}\hat{\vect{h}}_{2}^{\Htran}{\bf Z}^{-1}\hat{\vect{h}}_{1}|^2}{\big(\frac{1}{M} + \frac{1}{M} \hat{\bf h}_2^{\Htran}{\bf Z}^{-1}\hat{\vect{h}}_{2}\big)^2}
\end{align}
by applying again Lemma \ref{MIL} twice. 
Under Assumption~\ref{assumption_1} and by applying Lemma~\ref{lemma3}, 
\begin{align}
\frac{1}{M}\hat{\bf h}_1^{\Htran}{\bf Z}^{-2}\hat{\bf h}_1 &\asymp \frac{1}{M}\tr ( \vect{\Phi}_{1}{\bf Z}^{-2} )\triangleq \beta_{11}^{\prime}\\
\frac{1}{M}\hat{\bf h}_2^{\Htran}{\bf Z}^{-2}\hat{\bf h}_2 &\asymp\frac{1}{M}\tr ( \vect{\Phi}_{2}{\bf Z}^{-2} ) \triangleq \beta_{22}^{\prime}\\ 
\frac{1}{M}\hat{\bf h}_1^{\Htran}{\bf Z}^{-2}\hat{\bf h}_2 &\asymp\frac{1}{M}\tr (\vect{\Upsilon}_{12} {\bf Z}^{-2} ) \triangleq \beta_{12}^{\prime}
\end{align}
where $\beta_{11}^{\prime}$, $\beta_{22}^{\prime}$, and $\beta_{12}^{\prime}$ are non-negative real-valued scalars, since the trace of a product of positive semi-definite matrices is always non-negative.
Therefore, we obtain
\begin{align}\label{eq:C_10_1}
\frac{1}{M}\hat{\vect{h}}_{1}^{\Htran}\left(\hat{\vect{h}}_{2} \hat{\vect{h}}_{2}^{\Htran} + \vect{Z}  \right)^{-2} \hat{\vect{h}}_{1} &\asymp \beta_{11}^{\prime} - 2\frac{\beta_{12}\beta_{12}^{\prime}}{\beta_{22}}+\frac{\beta_{12}^2\beta_{12}^{\prime}}{(\beta_{22})^2} \triangleq \delta_1^{\prime} .
\end{align}
Plugging \eqref{eq:C_10_1} into \eqref{eq:C_10} and using $\frac{\gamma_{1}^{\rm {ul}}}{M} \asymp \delta_1$ yields $M\|\vect{v}_1\|^2 \asymp \frac{\delta_1^{\prime}}{\delta_1^2}$ such that 
\begin{align}
\frac{1}{\rho^{\rm dl}\vartheta_1}=\frac{{\mathbb{E}}\left\{\|\vect{v}_1\|^2\right\}}{\rho^{\rm dl}}\asymp \frac{1}{M\rho^{\rm dl}} \frac{\delta_1^{\prime}}{\delta_1^2 }. 
\end{align}

Consider now the two terms ${\mathbb{V}} \{ {\vect{h}}_{1}^{\Htran} {\vect{v}}_{1} \}$ and $\frac{\vartheta_2}{\vartheta_1}{\mathbb{E}}\big\{|{\vect{h}}_{1}^{\Htran} {\vect{v}}_{2}|^2\big\}$.  Similar to \cite[Eq.~(47)]{hoydis2013massive}, we can upper bound ${\mathbb{V}} \{ {\vect{h}}_{1}^{\Htran} {\vect{v}}_{1} \}$ as ${\mathbb{V}} \{ {\vect{h}}_{1}^{\Htran} {\vect{v}}_{1} \} \le 2{\mathbb{E}}\left\{\left|{\vect{h}}_{1}^{\Htran} {\vect{v}}_{1} - {\mathbb{E}}\left\{{\vect{h}}_{1}^{\Htran} {\vect{v}}_{1}\right\}\right|\right\} + {\mathbb{E}}\big\{\big|\tilde{\vect{h}}_{1}^{\Htran} {\vect{v}}_{1}\big|^2\big\}$. Notice that (by using ${\mathbb{E}}\left\{{\vect{h}}_{1}^{\Htran} {\vect{v}}_{1}\right\}\asymp1$ and the dominated convergence theorem) ${\mathbb{E}}\left\{\left|{\vect{h}}_{1}^{\Htran} {\vect{v}}_{1} - {\mathbb{E}}\left\{{\vect{h}}_{1}^{\Htran} {\vect{v}}_{1}\right\}\right|\right\} \asymp 0$ and
\begin{align}\notag
{\mathbb{E}}\big\{\big|\tilde{\vect{h}}_{1}^{\Htran} {\vect{v}}_{1}\big|^2\big\} &= {\mathbb{E}}\big\{{\vect{v}}_{1}^{\Htran}(\vect{R}_{1} - \vect{\Phi}_{1}){\vect{v}}_{1}\big\}\\&\mathop \leq\limits^{(a)}\|\vect{R}_{1} - \vect{\Phi}_{1} \|_2{\mathbb{E}}\big\{\|\vect{v}_1\|^2\big\}\mathop \asymp\limits^{(b)}0
\end{align}
where $(a)$ and $(b)$ follow from Lemma \ref{lemma2} and ${\mathbb{E}}\big\{\|\vect{v}_1\|^2\big\} \asymp 0$ (since, as shown above, $\|\vect{v}_1\|^2 \asymp \frac{1}{M}\frac{\delta_1^{\prime}}{\delta_1^2} \asymp 0 $), respectively. Therefore, we have that ${\mathbb{V}} \{ {\vect{h}}_{1}^{\Htran} {\vect{v}}_{1} \} \asymp 0$. Finally, we consider $\frac{\vartheta_2}{\vartheta_1}{\mathbb{E}}\big\{|{\vect{h}}_{1}^{\Htran} {\vect{v}}_{2}|^2\big\}$. By using \eqref{eq:beta_22}, \eqref{eq:beta_12}, and $\liminf_M\beta_{11}>0$ (as follows from Assumption~\ref{assumption_1}), we have that 
\begin{align}\notag
{\vect{h}}_{1}^{\Htran} {\vect{v}}_{2}&\mathop =\limits^{(a)} \frac{{\vect{h}}_{1}^{\Htran} \left(\hat{\vect{h}}_{1} \hat{\vect{h}}_{1}^{\Htran} + \vect{Z}  \right)^{-1} \hat{\vect{h}}_{2}}{1+\hat{\vect{h}}_{2}^{\Htran}\left(\hat{\vect{h}}_{1} \hat{\vect{h}}_{1}^{\Htran} + \vect{Z}  \right)^{-1}\hat{\vect{h}}_{2}} \\& \mathop =\limits^{(b)} \frac{\frac{1}{M}{\vect{h}}_{1}^{\Htran}\vect{Z}^{-1}\hat{\vect{h}}_{2} - \frac{\frac{1}{M}{\vect{h}}_{1}^{\Htran}\vect{Z}^{-1}\hat{\vect{h}}_{1}\frac{1}{M}\hat{\vect{h}}_{1}\vect{Z}^{-1}\hat{\vect{h}}_{2}^{\Htran}}{\frac{1}{M}+\frac{1}{M}\hat{\vect{h}}_{1}^{\Htran}\vect{Z}^{-1}\hat{\vect{h}}_{1}}}{\frac{1}{M}+\frac{1}{M}\gamma_{2}^{\rm {ul}}}\notag\\&\mathop \asymp\limits^{(c)} \frac{\beta_{12} - \frac{\beta_{11}\beta_{12}}{\beta_{11}}}{\delta_2} = 0
\end{align}
where $(a)$ and $(b)$ follow from Lemma \ref{MIL} after identifying\footnote{The uplink SINR $\gamma_{2}^{\rm {ul}}$ of UE~2 is obtained from \eqref{eq:gamma1_MMSE} by interchanging UE indices.} $\hat{\vect{h}}_{2}^{\Htran}\big(\hat{\vect{h}}_{1} \hat{\vect{h}}_{1}^{\Htran} + \vect{Z}  \big)^{-1}\hat{\vect{h}}_{2}$ as $\gamma_{2}^{\rm {ul}}$ (by also dividing and multiplying by $M$), and $(c)$ follows by using \eqref{eq:beta_11}, \eqref{eq:beta_12} and the fact that 
\begin{align}
\frac{\gamma_{2}^{\rm {ul}}}{M} \asymp \delta_2 \triangleq \beta_{22} - \frac{\beta_{21}^2}{ \beta_{11}}
\label{eq:asympotitic_SINR_2}
\end{align}
with $\liminf_M \delta_2 >0$ (which follows from the proof of Theorem~\ref{theorem:MMSE}  by interchanging UE indices).
By applying Lemma \ref{lemma3}, this implies ${\mathbb{E}}\big\{|{\vect{h}}_{1}^{\Htran} {\vect{v}}_{2}|^2\big\}\asymp 0$. {Observe now that $\frac{\vartheta_2}{\vartheta_1} \asymp \frac{\delta_{1}^\prime}{\delta_1^{2}}\frac{\delta_2^{2}}{\delta_{2}^\prime}$ where $\delta_{2}^\prime$ is obtained from $\delta_{1}^\prime$ by interchanging UE indices. Since all the quantities in $\delta_1^{\prime}$ are uniformly bounded (due to Assumption~\ref{assumption_1}), $\liminf_M\delta_1>0$ (as proved in Appendix~B) and $\liminf_M\delta_2<\infty$ (since from \eqref{eq:asympotitic_SINR_2} $\delta_2 < \beta_{22}$ and $\liminf_M \beta_{22} < \infty$ due to Assumption~\ref{assumption_1}), we eventually have that $\frac{\vartheta_2}{\vartheta_1}{\mathbb{E}}\big\{|{\vect{h}}_{1}^{\Htran} {\vect{v}}_{2}|^2\big\}\asymp 0$. }

Combining all the above results yields 
\begin{align} \label{eq:C10_DL}
\frac{\gamma_{1}^{\rm {dl}}}{M}  \asymp  \rho^{\rm {dl}} \frac{\delta_1^2}{\delta_1^{\prime}}.
\end{align}
Since all the quantities in $\delta_1^{\prime}$ are uniformly bounded and $\liminf_M\delta_1>0$, it follows that $\gamma_{1}^{\rm {dl}}$ grows unboundedly as $M \to \infty$.
This implies that also $\mathsf{SE}_{1}^{\rm {dl}}$ grows unboundedly as $M\to \infty$, which can be proved by the same arguments as in the last paragraph of Appendix~B.

\section*{Appendix E -- Proof of Theorem~\ref{theorem:EW-MMSE_precoding}}\label{appendix:proof-theorem:EW_MMSE_precoding}
The EW-MMSE estimate $\hat{\vect{h}}_{k} $ and the estimation error $\tilde{\vect{h}}_{k}= \vect{h}_{k} - \hat{\vect{h}}_{k}$ are random vectors distributed as $\hat{\vect{h}}_{k}  \sim \CN({\bf 0},{\boldsymbol \Sigma}_k)$ and $\tilde{\vect{h}}_{k}  \sim \CN({\bf 0},{\tilde{\boldsymbol \Sigma}}_k)$ with ${\boldsymbol \Sigma}_k = \vect{D}_{k}
\boldsymbol{\Lambda}^{-1}\vect{Q}\boldsymbol{\Lambda}^{-1}\vect{D}_{k}$ and $\tilde {\boldsymbol \Sigma}_k = \vect{R}_{k} - \vect{D}_{k}
\boldsymbol{\Lambda}^{-1}\vect{R}_{k} - \vect{R}_{k}
\boldsymbol{\Lambda}^{-1}\vect{D}_{k} - {\boldsymbol \Sigma}_k$. Unlike with MMSE estimation, the vectors $\hat{\vect{h}}_{k} $ and $\tilde{\vect{h}}_{k}$ are correlated with ${\mathbb{E}}\{\hat{\vect{h}}_{k}\tilde{\vect{h}}_{k}^{\Htran}\} = {\mathbb{E}}\{\hat{\vect{h}}_{k}({\vect{h}}_{k} - \hat{\vect{h}}_{k})^{\Htran}\}= \vect{D}_{k}
\boldsymbol{\Lambda}^{-1}\vect{R}_{k} - {\boldsymbol \Sigma}_k$. 
Hence, ${\vect{v}}_{1}$ and $\tilde{\vect{h}}_{1}$ are also correlated.
For later convenience, we also notice that ${\mathbb{E}}\{{\vect{h}}_{1}\hat{\vect{h}}_{1}^{\Htran}\} = \vect{R}_1\boldsymbol{\Lambda}^{-1}\vect{D}_{1} $, ${\mathbb{E}}\{{\vect{h}}_{1}\hat{\vect{h}}_{2}^{\Htran}\} = \vect{R}_1\boldsymbol{\Lambda}^{-1}\vect{D}_{2}$, and ${\mathbb{E}}\{\hat{\vect{h}}_{2}\hat{\vect{h}}_{1}^{\Htran}\} = \vect{D}_{2}
\boldsymbol{\Lambda}^{-1}\vect{Q}\boldsymbol{\Lambda}^{-1}\vect{D}_{1} = \vect{\Theta}_{21}$.
We need to characterize all the terms in \eqref{eq:G_102} and begin with ${\mathbb{E}}\left\{\vect{v}_{1}^{\Htran} {\vect{h}}_{1}\right\}$. By applying Lemma \ref{MIL} and by dividing and multiplying by $M$, we can express ${\vect{v}}_{1}^{\Htran}{\vect{h}}_{1}$ as
\begin{align}\label{eq:G.1}
{\vect{v}}_{1}^{\Htran}{\vect{h}}_{1} = \frac{\frac{1}{M}{\hat{\vect{h}}}_{1}^{\Htran}\left(\hat{\vect{h}}_{2} \hat{\vect{h}}_{2}^{\Htran} + \vect{S}  \right)^{-1}{\vect{h}}_{1}}{{\frac{1}{M} + \frac{1}{M}{\hat{\vect{h}}}_{1}^{\Htran}\left(\hat{\vect{h}}_{2} \hat{\vect{h}}_{2}^{\Htran} + \vect{S}  \right)^{-1} \hat{\vect{h}}_{1}}} = \frac{\frac{1}{M}\tilde \mu_{1}^{\rm ul}}{\frac{1}{M} + \frac{1}{M}{\mu}_{1}^{\rm ul}}.
\end{align}
Notice that ${\mu}_{1}^{\rm ul}$ has the same form as $\gamma_{1}^{\rm {ul}}$ in \eqref{eq:gamma1_MMSE}, but with $\{\hat{\vect{h}}_{k}:  k=1,2\}$ now given by \eqref{eq:G_101}. Under Assumption~\ref{assumption_1} and by Lemma~\ref{lemma3},\footnote{The expressions in \eqref{eq:alpha_11}--\eqref{eq:alpha_12} have been simplified by utilizing the fact that $\vect{Q}$ and $\vect{\Lambda}$ have the same diagonal elements and $\vect{R}_k$ and $\vect{D}_k$ have the same diagonal elements, for $k=1,2$.}
\begin{align}\label{eq:alpha_11}
\frac{1}{M}\hat{\bf h}_1^{\Htran}\vect{S}^{-1}\hat{\bf h}_1 &\asymp \frac{1}{M}\tr ( \vect{\Sigma}_{1} \vect{S}^{-1}) = \frac{1}{M}\sum_{i=1}^M\frac{[\vect{R}_{1}]_{i,i}^2}{[\vect{S}]_{i,i}[\vect{\Lambda}]_{i,i}} \triangleq \alpha_{11}\\\label{eq:alpha_22}
\frac{1}{M}\hat{\bf h}_2^{\Htran}\vect{S}^{-1}\hat{\bf h}_2 &\asymp\frac{1}{M}\tr ( \vect{\Sigma}_{2} \vect{S}^{-1}) = \frac{1}{M}\sum_{i=1}^M\frac{[\vect{R}_{2}]_{i,i}^2}{[\vect{S}]_{i,i}[\vect{\Lambda}]_{i,i}} \triangleq \alpha_{22}\\ 
\frac{1}{M}\hat{\bf h}_1^{\Htran}\vect{S}^{-1}\hat{\bf h}_2 &\asymp\frac{1}{M}\tr (\vect{\Theta}_{21} \vect{S}^{-1}) \notag\\ &= \frac{1}{M}\sum_{i=1}^M\frac{[\vect{R}_{1}]_{i,i}[\vect{R}_{2}]_{i,i}}{[\vect{S}]_{i,i}[\vect{\Lambda}]_{i,i}} \triangleq \alpha_{12}. \label{eq:alpha_12}
\end{align}
By applying the same line of reasoning as when analyzing $\gamma_{1}^{\rm {ul}}$ in Appendix B and exploiting the fact that $\liminf_M\alpha_{22}>0$ (which follows from Assumption~\ref{assumption_1}), we obtain $\frac{\mu_{1}^{\rm ul}}{M} = \frac{1}{M}{\hat{\vect{h}}}_{1}^{\Htran}\left(\hat{\vect{h}}_{2} \hat{\vect{h}}_{2}^{\Htran} + \vect{S}  \right)^{-1}{\vect{h}}_{1}\asymp \upsilon_1 \triangleq \alpha_{11} - \frac{\alpha_{12}^2}{\alpha_{22}}$.  Note that $\liminf_M \upsilon_1 > 0$ under Assumption~\ref{assumption_6}. This can be proved, as done in Appendix B for $\delta_1$, by expanding the condition reported in the corollary below (the proof unfolds from the same arguments as in Appendix~C).
\begin{corollary}\label{corollary_6} If Assumption~\ref{assumption_6} holds, then for $\boldsymbol{\lambda} = [\lambda_1, \lambda_2]^{\Ttran} \in \mathbb{R}^2$ and $i=1,2$,
\begin{align} \notag
& \mathop {\liminf}\limits_M \inf_{\{\boldsymbol{\lambda}: \, \lambda_i=1\}} \\ &\frac{1}{{M}}\tr \Big( \boldsymbol{\Lambda}^{-1}\vect{Q}\boldsymbol{\Lambda}^{-1}\big(\lambda_1\vect{D}_{1} +\lambda_2 \vect{D}_{2}\big)\vect{S}^{-1}\big(\lambda_1\vect{D}_{1} +\lambda_2 \vect{D}_{2}\big)  \Big) > 0. \label{eq:assumption_6_expanded}
\end{align}
\end{corollary} 

As for $\tilde \mu_{1}^{\rm ul}$ in \eqref{eq:G.1}, we have that
\begin{align}\notag
\frac{1}{M}\tilde \mu_{1}^{\rm ul} &= \frac{1}{M}{\hat{\vect{h}}}_{1}^{\Htran}\left(\hat{\vect{h}}_{2} \hat{\vect{h}}_{2}^{\Htran} + \vect{S}  \right)^{-1}{\vect{h}}_{1} \\&= \frac{1}{M} {\hat {\vect{h}}}_{1}^{\Htran}\vect{S}^{-1}{\vect{h}}_{1} - \frac{\frac{1}{M}{\hat {\vect{h}}}_{1}^{\Htran}\vect{S}^{-1}\hat{\vect{h}}_{2}\frac{1}{M}\hat{\vect{h}}_{2}^{\Htran}\vect{S}^{-1}{\vect{h}}_{1}}{\frac{1}{M}+\frac{1}{M}\hat{\vect{h}}_{2}^{\Htran}\vect{S}^{-1}\hat{\vect{h}}_{2}} \asymp  \upsilon_1
\end{align}
since the diagonal structure of the matrices $\boldsymbol{\Lambda}$, $\vect{D}_{1}$, $\vect{D}_{2}$, and $\vect{S}$ implies that 
\begin{align}\notag
\frac{1}{M}\hat{\bf h}_1^{\Htran}\vect{S}^{-1}{\bf h}_1 &\asymp \frac{1}{M}\tr ( \vect{R}_{1}\boldsymbol{\Lambda}^{-1}\vect{D}_{1}\vect{S}^{-1}) \\&= \frac{1}{M}\sum_{i=1}^M\frac{[\vect{R}_{1}]_{i,i}^2}{[\vect{S}]_{i,i}[\vect{\Lambda}]_{i,i}} = \alpha_{11}\\\notag
\frac{1}{M}\hat{\bf h}_2^{\Htran}\vect{S}^{-1} {\bf h}_1 &\asymp\frac{1}{M}\tr (\vect{R}_{1} \boldsymbol{\Lambda}^{-1}\vect{D}_{2} \vect{S}^{-1}) \\&= \frac{1}{M}\sum_{i=1}^M\frac{[\vect{R}_{1}]_{i,i}[\vect{R}_{2}]_{i,i}}{[\vect{S}]_{i,i}[\vect{\Lambda}]_{i,i}}  = \alpha_{12}.
\end{align}
Applying these results to \eqref{eq:G.1} yields $\vect{v}_{1}^{\Htran} {\vect{h}}_{1} \asymp 1$ from which it follows that $|{\mathbb{E}}\{\vect{v}_{1}^{\Htran} {{\vect{h}}}_{1}\}|^2 \asymp 1$. 

\begin{figure*}
\begin{align}\label{eq:sec_E_inverse_of_MMSE_matrix}
\left(   {\bf A}_{j,\setminus k}+ \hat{\vect{H}}_{jk,\setminus j}\hat{\vect{H}}_{jk,\setminus j}^{\Htran}\right)^{-1} = {\bf A}_{j,\setminus k}^{-1} -   {\bf A}_{j,\setminus k}^{-1}\hat{\vect{H}}_{jk,\setminus j}\left(   {\bf I}_{L-1}+ \hat{\vect{H}}_{jk,\setminus j}^{\Htran}{\bf A}_{j,\setminus k}^{-1}\hat{\vect{H}}_{jk,\setminus j}\right)^{-1}\!\hat{\vect{H}}_{jk,\setminus j}^{\Htran}{\bf A}_{j,\setminus k}^{-1}.
\end{align}
\hrule
\begin{align} \frac{\gamma_{jk}^{\rm {ul}}}{M} =  \frac{1}{M} \hat{\vect{h}}_{jjk}^{\Htran}{\bf A}_{j,\setminus k}^{-1}\hat{\vect{h}}_{jjk}  - 
 \frac{1}{M} \hat{\vect{h}}_{jjk}^{\Htran}  {\bf A}_{j,\setminus k}^{-1}\hat{\vect{H}}_{jk,\setminus j}\left(  \frac{1}{M} {\bf I}_{L-1}+ \frac{1}{M}\hat{\vect{H}}_{jk,\setminus j}^{\Htran}{\bf A}_{j,\setminus k}^{-1}\hat{\vect{H}}_{jk,\setminus j}\right)^{-1}\!\!\!\frac{1}{M}\hat{\vect{H}}_{jk,\setminus j}^{\Htran}{\bf A}_{j,\setminus k}^{-1}\hat{\vect{h}}_{jjk}.\label{eq:appendixc_7}
\end{align}
\hrule
\begin{align} \notag
\frac{1}{M}{\bf A}_{j,\setminus k}^{-1} & = \frac{1}{M}\Big(\sum\limits_{l} \sum\limits_{i\ne k} \hat{\vect{h}}_{jli}\hat{\vect{h}}_{jli}^{\Htran} + \vect{Z}_j\Big)^{-1} = \frac{1}{M}\Big( \hat{\vect{H}}_{j,\setminus k} \hat{\vect{H}}_{j,\setminus k}^{\Htran} + \vect{Z}_j\Big)^{-1} \\ &= \frac{1}{M}\vect{Z}_j^{-1} - \frac{1}{M}\vect{Z}_j^{-1}\hat{\vect{H}}_{j,\setminus k} \left(  \frac{1}{M} {\bf I}_{L(K-1)}+ \frac{1}{M}\hat{\vect{H}}_{j,\setminus k}^{\Htran}\vect{Z}_j^{-1}\hat{\vect{H}}_{j,\setminus k}\right)^{-1}\!\!\!\frac{1}{M}\hat{\vect{H}}_{j,\setminus k}^{\Htran}\vect{Z}_j^{-1}\label{eq:newInFigure}
\end{align}
\hrule
\end{figure*}

Next, consider the noise term $\frac{1}{\rho^{\rm ul}}  {\mathbb{E}}\left\{||\vect{v}_{1} ||^2\right\}$ for which 
\begin{align}
\|\vect{v}_1\|^2 &= \frac{\hat{\vect{h}}_{1}^{\Htran}\left(\hat{\vect{h}}_{2} \hat{\vect{h}}_{2}^{\Htran} + \vect{S} \right)^{-2} \hat{\vect{h}}_{1} }{\left(1 + \mu_1^{\rm ul}\right)^2} \\ &= \frac{1}{M} \frac{\frac{1}{M} \hat{\vect{h}}_{1}^{\Htran}\left(\hat{\vect{h}}_{2} \hat{\vect{h}}_{2}^{\Htran} + \vect{S}\right)^{-2} \hat{\vect{h}}_{1} }{\left(\frac{1}{M} + \frac{1}{M} \mu_{1}^{\rm ul}\right)^2}  . \label{eq:G_10}
\end{align}
Under Assumption~\ref{assumption_1} and by  Lemma~\ref{lemma3}, 
\begin{align}
\frac{1}{M}\hat{\bf h}_1^{\Htran}{\bf S}^{-2}\hat{\bf h}_1 &\asymp \frac{1}{M}\tr ( \vect{\Sigma}_{1}{\bf S}^{-2} ) = \frac{1}{M}\sum_{i=1}^M\frac{[\vect{R}_{1}]_{i,i}^2}{[\vect{S}]_{i,i}^2[\vect{\Lambda}]_{i,i}}\triangleq \alpha_{11}^{\prime}\\
\frac{1}{M}\hat{\bf h}_2^{\Htran}{\bf S}^{-2}\hat{\bf h}_2 &\asymp\frac{1}{M}\tr ( \vect{\Sigma}_{2}{\bf S}^{-2} )  = \frac{1}{M}\sum_{i=1}^M\frac{[\vect{R}_{2}]_{i,i}^2}{[\vect{S}]_{i,i}^2[\vect{\Lambda}]_{i,i}} \triangleq \alpha_{22}^{\prime}\\ \notag
\frac{1}{M}\hat{\bf h}_1^{\Htran}{\bf S}^{-2}\hat{\bf h}_2 &\asymp\frac{1}{M}\tr (\vect{\Theta}_{21} {\bf S}^{-2} ) \\ &= \frac{1}{M}\sum_{i=1}^M\frac{[\vect{R}_{1}]_{i,i}[\vect{R}_{2}]_{i,i}}{[\vect{S}]_{i,i}^2[\vect{\Lambda}]_{i,i}}\triangleq \alpha_{12}^{\prime}
\end{align}
where $\alpha_{11}^{\prime}$, $\alpha_{22}^{\prime}$, and $\alpha_{12}^{\prime}$ are non-negative real-valued scalars. 
{By applying Lemma \ref{MIL} twice to the numerator in \eqref{eq:G_10} and by using the above results, we obtain}
\begin{align}\label{eq:G_10_1}
\frac{1}{M} \hat{\vect{h}}_{1}^{\Htran}\left(\hat{\vect{h}}_{2} \hat{\vect{h}}_{2}^{\Htran} + \vect{S} \right)^{-2} \hat{\vect{h}}_{1}  \asymp  \alpha_{11}^{\prime} - 2\frac{\alpha_{12}\alpha_{12}^{\prime}}{\alpha_{22}}+\frac{\alpha_{12}^2\alpha_{22}^{\prime}}{(\alpha_{22})^2} \triangleq \upsilon_1^\prime.
\end{align}
Plugging \eqref{eq:G_10_1} into \eqref{eq:G_10} yields $M\|\vect{v}_1\|^2 \asymp \frac{\upsilon_1^{\prime}}{\upsilon_{1}^2}$ such that 
$
\frac{1}{\rho^{\rm ul}}  {\mathbb{E}}\left\{||\vect{v}_{1} ||^2\right\}\asymp  \frac{1}{M\rho^{\rm ul}} \frac{\upsilon_1^{\prime}}{\upsilon_1^2}. 
$

As for ${\mathbb{V}} \{ {\vect{v}}_{1}^{\Htran} {\vect{h}}_{1} \}$, it can be easily proved (using the above results and Lemma \ref{lemma3}), that ${\mathbb{V}} \{ {\vect{v}}_{1}^{\Htran} {\vect{h}}_{1} \} \asymp 0$.
Consider now the interference term ${\mathbb{E}}\big\{| {\vect{v}}_{1}^{\Htran}{\vect{h}}_{2}|^2\big\}$. Using \eqref{eq:alpha_22} and \eqref{eq:alpha_12}, we have (by applying Lemma \ref{MIL} and dividing and multiplying by $M$) that
 \begin{align}\notag
{\vect{v}}_{1}^{\Htran}{\vect{h}}_{2}&= \frac{\frac{1}{M}\hat {\vect{h}}_{1}^{\Htran} \left(\hat{\vect{h}}_{2} \hat{\vect{h}}_{2}^{\Htran} +{\bf S} \right)^{-1} {\vect{h}}_{2}}{\frac{1}{M}+\frac{1}{M}\mu_1^{\rm {ul}}} \\&= \frac{\frac{1}{M}\hat{\vect{h}}_{1}^{\Htran}{\bf S}^{-1}{\vect{h}}_{2} - \frac{ \frac{1}{M} \hat{\vect{h}}_{1}^{\Htran}{\bf S}^{-1}\hat{\vect{h}}_{2}\frac{1}{M}\hat{\vect{h}}_{2}^{\Htran}{\bf S}^{-1}{\vect{h}}_{2}}{\frac{1}{M}+\frac{1}{M}\hat{\vect{h}}_{2}^{\Htran}{\bf S}^{-1}\hat{\vect{h}}_{2}}}{\frac{1}{M}+\frac{1}{M}\mu_1^{\rm {ul}}}\notag\\&\asymp \frac{\alpha_{12} - \frac{\alpha_{12}\alpha_{22}}{\alpha_{22}}}{\upsilon_1} = 0\label{eq:appendixF_interference}
\end{align}
where we have used the fact that $\frac{1}{M}\hat{\vect{h}}_{1}^{\Htran}{\bf S}^{-1}{\vect{h}}_{2} \asymp \alpha_{12}$ and $
\frac{1}{M}\hat{\vect{h}}_{2}^{\Htran}{\bf S}^{-1}{\vect{h}}_{2}\asymp \alpha_{22}$.
Applying Lemma \ref{lemma3} to \eqref{eq:appendixF_interference}, we obtain ${\mathbb{E}}\big\{|{\vect{v}}_{1}^{\Htran} {\vect{h}}_{2}|^2\big\}\asymp0$.

Combining all the above results together yields $\frac{{\underline{\gamma}}_{1}^{\rm ul}}{M} \asymp  \rho^{\rm{ul}}\frac{\upsilon_1^2}{\upsilon_1^{\prime}}$.
Since all the components of $\upsilon_1^{\prime}$ in \eqref{eq:G_10_1} are uniformly bounded and $\liminf_M\upsilon_1>0$ (under Assumption~\ref{assumption_6}), it follows that ${\underline{\gamma}}_{1}^{\rm ul}$ grows unboundedly as $M \to \infty$. Hence, ${\underline{\mathsf{SE}}}_{1}^{\rm ul}$ also grows without bound.

\section*{Appendix F -- Proof of Theorem \ref{theorem:M-MMSE}}\label{proof:Theorem4}
We start by rewriting $\gamma_{jk}^{\rm{ul}}$ in \eqref{eq:gammajk_MMSE} as
\begin{align}\label{eq:appendixc_1}
\gamma_{jk}^{\rm {ul}} =  
 \hat{\vect{h}}_{jjk}^{\Htran}  \Bigg(  \underbrace{\sum_{l} \sum_{i\ne k} \hat{\vect{h}}_{jli}\hat{\vect{h}}_{jli}^{\Htran} +   \vect{Z}_j}_{{\bf A}_{j,\setminus k}} + \underbrace{\sum\limits_{l\ne j}  \hat{\vect{h}}_{jlk}\hat{\vect{h}}_{jlk}^{\Htran}}_{\hat{\vect{H}}_{jk,\setminus j}\hat{\vect{H}}_{jk,\setminus j}^{\Htran}}\Bigg)^{-1} \hat{\vect{h}}_{jjk} 
\end{align}
where ${\bf A}_{j,\setminus k} = \sum_{l} \sum_{i\ne k} \hat{\vect{h}}_{jli}\hat{\vect{h}}_{jli}^{\Htran} +   \vect{Z}_j$ is
independent of $\{\hat{\vect{h}}_{jlk}: l=1,\ldots L\} $ 	and $\hat{\vect{H}}_{jk,\setminus j} = [\hat{\vect{h}}_{j1k} \ldots \hat{\vect{h}}_{jj-1k} \,  \hat{\vect{h}}_{jj+1k} \ldots \hat{\vect{h}}_{jLk}] \in\mathbb{C}^{M\times(L-1)}$ collects all vectors $\hat{\vect{h}}_{jlk}$  with $l\ne j$ (i.e., the channels of UEs that cause pilot contamination). By Lemma \ref{MIL}, we obtain \eqref{eq:sec_E_inverse_of_MMSE_matrix} at the top of the page.
Plugging \eqref{eq:sec_E_inverse_of_MMSE_matrix} into \eqref{eq:appendixc_1} and dividing both sides by $M$ leads to \eqref{eq:appendixc_7}.
By applying Lemma \ref{MIL} once again, \eqref{eq:newInFigure} follows 
where $\hat{\vect{H}}_{j,\setminus k}\in\mathbb{C}^{M\times L(K-1)}$ denotes the matrix collecting all vectors $\hat{\vect{h}}_{jli}$  with $i\ne k$, which is independent of $\hat{\vect{h}}_{jlk}$ for any $j$ and $l$. Therefore, it follows that the first term in \eqref{eq:appendixc_7} is such that
\begin{align}\notag
\frac{1}{M}\hat{\vect{h}}_{jjk}^{\Htran}{\bf A}_{j,\setminus k}^{-1}\hat{\vect{h}}_{jjk}   & \mathop\asymp^{{(a)}} \frac{1}{M}\hat{\vect{h}}_{jjk}^{\Htran}\vect{Z}_j^{-1}\hat{\vect{h}}_{jjk} \\&\mathop\asymp^{{(b)}} \frac{1}{M}\tr  ( \vect{\Phi}_{jjk} {\vect{Z}}_{j}^{-1} ) \triangleq \beta_{jk,jj}
 \end{align}
where ${{(a)}}$ follows from Lemma~\ref{lemma3} since $\hat{\vect{h}}_{jjk}$ and  $\hat{\vect{H}}_{j,\setminus k}$ are independent and thus $\frac{1}{M}\hat{\vect{h}}_{jjk}^{\Htran}\vect{Z}^{-1}\hat{\vect{H}}_{j,\setminus k} \asymp {\vect{0}}_{L(K-1)}$ (remember that $\hat{\vect{H}}_{j,\setminus k}$ collects the $L(K-1)$ vectors $\{\hat{\vect{h}}_{jli}\}$ with $i\ne k$), and ${{(b)}}$ follows from Lemma \ref{lemma3} by recalling that $\hat{\vect{h}}_{jjk}\sim  \CN \left( \vect{0},  \vect{\Phi}_{jjk} \right)$ where the matrices $\vect{\Phi}_{jjk} $ can be proved (using Lemma~\ref{lemma2}) to have uniformly bounded spectral norm due to Assumption \ref{assumption_3}. Using similar arguments, we have that the $l$th element of the row vector $\frac{1}{M}\hat{\vect{h}}_{jjk}^{\Htran}  {\bf A}_{j,\setminus k}^{-1}\hat{\vect{H}}_{jk,\setminus j} \in \mathbb{C}^{1\times (L-1)}$ is such that
 \begin{align}\notag
  \left[\frac{1}{M}\hat{\vect{h}}_{jjk}^{\Htran}  {\bf A}_{j,\setminus k}^{-1}\hat{\vect{H}}_{jk,\setminus j}\right]_l  & \asymp \frac{1}{M} \hat{\vect{h}}_{jjk}^{\Htran}  {\bf Z}_j^{-1}\hat{\vect{h}}_{jlk} \\&\hspace{-3cm}\asymp \frac{1}{M}\tr (\vect{R}_{jlk} \vect{Q}_{jk}^{-1} \vect{R}_{jjk}{\vect{Z}}_{j}^{-1} )  \triangleq \beta_{jk,lj}\label{eq:Section_Appendix_81}
   \end{align}
   for $l=1,2,\ldots,L-1$.
Furthermore, the $(n,m)$th element of $\frac{1}{M} \hat{\vect{H}}_{jk,\setminus j}^{\Htran}{\bf A}_{j,\setminus k}^{-1}\hat{\vect{H}}_{jk,\setminus j}$ is
    \begin{align}\notag
   \frac{1}{M}\left[\hat{\vect{H}}_{jk,\setminus j}^{\Htran}{\bf A}_{j,\setminus k}^{-1}\hat{\vect{H}}_{jk,\setminus j}\right]_{n,m} &\asymp \frac{1}{M}\hat{\vect{h}}_{jnk}^{\Htran}\vect{Z}_j^{-1}\hat{\vect{h}}_{jmk} \\&\hspace{-3.5cm}\asymp \frac{1}{M}\tr  ( \vect{R}_{jmk} \vect{Q}_{jk}^{-1} \vect{R}_{jnk}{\vect{Z}}_{j}^{-1})  \triangleq \beta_{jk,mn}.\label{eq:Section_Appendix_82}
     \end{align} 
    For notational convenience, let us define $\vect{b}_{jk}\in \mathbb{R}^{L-1}$ and $\vect{C}_{jk} \in \mathbb{R}^{(L-1)\times (L-1)}$ with entries
\begin{align}\notag
\big[{\vect{b}_{jk}}\big]_l = \beta_{jk,lj} &= \vecoperator \left(\frac{1}{\sqrt M} {\vect{Z}}_{j}^{-1/2}\vect{R}_{jlk} \vect{Q}_{jk}^{-1/2}\right)^{\Htran}\\&\vecoperator  \left(\frac{1}{\sqrt M}{\vect{Z}}_{j}^{-1/2}\vect{R}_{jjk} \vect{Q}_{jk}^{-1/2}\right)
\end{align}
and
\begin{align}\notag
\big[\vect{C}_{jk} \big]_{l,n} =  \beta_{jk,ln}  &= \vecoperator \left(\frac{1}{\sqrt M}{\vect{Z}}_{j}^{-1/2}\vect{R}_{jlk} \vect{Q}_{jk}^{-1/2}\right)^{\Htran}\\&\vecoperator \left(\frac{1}{\sqrt M}{\vect{Z}}_{j}^{-1/2}\vect{R}_{jnk} \vect{Q}_{jk}^{-1/2}\right) \label{eq:C_jk}
\end{align}
where we have used the fact that $\tr({\bf AB}) = \vecoperator({\bf A}^{\Htran})^{\Htran}\vecoperator({\bf B})$. 
In Appendix~G, it is shown that, under Assumption~\ref{assumption_4}, the following corollary holds.
{\begin{corollary}\label{assumption_4_1} If Assumption~\ref{assumption_4} holds, then for any UE $k$ in cell $j$ with ${\boldsymbol\lambda}_{jk} = [\lambda_{j1k}, \ldots,\lambda_{jLk}]^{\Ttran} \in \mathbb{R}^{L}$ and $l'=1,\ldots,L$ 
\begin{align} \notag 
	& \liminf_M \inf_{\{{\boldsymbol\lambda}_{jk}:\lambda_{jl'k}=1\}} \\  &\frac{1}{{M}}\tr \Bigg( \vect{Q}_{jk}^{-1} \Big( \sum\limits_{l=1}^L\lambda_{jlk} \vect{R}_{jlk}\Big) \vect{Z}_{j}^{-1}  \Big(\sum\limits_{l=1}^L\lambda_{jlk} \vect{R}_{jlk}\Big)  \Bigg) > 0 \label{Condition2_Assumption4_11}
	\end{align}
	and the matrix $\vect{C}_{jk}$ is invertible as $M\to \infty$.
\end{corollary} 
Since $\vect{C}_{jk}$ is invertible as $M\to \infty$ under Assumption~\ref{assumption_4}, we have that $\frac{\gamma_{jk}^{\rm {ul}}}{M}$ in \eqref{eq:appendixc_7} is such that\begin{align}\label{eq:general result}
\frac{\gamma_{jk}^{\rm {ul}}}{M} \asymp \delta_{jk} \triangleq \beta_{jj,jk} - \vect{b}_{jk}^{\Htran}\vect{C}_{jk}^{-1}{\vect{b}_{jk}}.
\end{align}
Expanding condition {\eqref{Condition2_Assumption4_11}} in Corollary~\ref{assumption_4_1} for $\lalt=j$ and using the definitions of $\vect{b}_{jk}$ and $\vect{C}_{jk}$ yield
\begin{align}\label{eq:Appendix_E.1_3}
\liminf_M \inf_{\overline{\boldsymbol\lambda}_{jk}} \left(\beta_{jj,jk} +2\overline{\boldsymbol\lambda}_{jk}^{\Ttran}\vect{b}_{jk}+  \overline{\boldsymbol\lambda}_{jk}^{\Ttran}\vect{C}_{jk}\overline{\boldsymbol\lambda}_{jk}\right) >0
	\end{align}
	with $\overline{\boldsymbol\lambda}_{jk} =[\lambda_{j1k}, \ldots,\lambda_{j(j-1)k}, \lambda_{j(j+1)k}, \ldots,\lambda_{jLk}]^{\Ttran} \in \mathbb{R}^{L-1}$.
The invertibility of $\vect{C}_{jk}$ as $M\to \infty$ ensures that the infimum exists for sufficiently large $M$ and that it is given by
\begin{align} \notag
 \inf_{\overline{\boldsymbol{\lambda}}_{jk}}\big(\beta_{jj,jk} + 2\overline{\boldsymbol{\lambda}}_{jk}^{\Ttran}\vect{b}_{jk}&+  \overline{\boldsymbol{\lambda}}_{jk}^{\Ttran}\vect{C}_{jk}\overline{\boldsymbol{\lambda}}_{jk}\big) \\&= \beta_{jj,jk} - {\bf b}_{jk}^{\Ttran}\vect{C}_{jk}^{-1}{\bf b}_{jk} = \delta_{jk}\label{eq:Appendix_E.1_4}
\end{align}
where the infimum is attained by $\overline{\boldsymbol{\lambda}}_{jk} = \vect{C}_{jk}^{-1}{\bf b}_{jk}$. Substituting \eqref{eq:Appendix_E.1_4} into \eqref{eq:Appendix_E.1_3} implies that $\liminf_M\delta_{jk}>0$. Therefore, $\gamma_{jk}^{\rm {ul}}$ grows a.s.~unboundedly and this implies that $\mathsf{SE}_{jk}^{\rm {ul}}$ grows unboundedly as $M\to \infty$, which can be proved as done in the last paragraph of Appendix~B.}

{
\section*{Appendix G -- Proof of Corollary~\ref{assumption_4_1} in Appendix F} \label{appendix:invertibility-C}
The argument of the left-hand side of \eqref{Condition2_Assumption4_11} can be lower bounded by
\begin{align}\label{eq:AppendixG_99}
\frac{\frac{1}{{M}}\big\| \sum_{l=1}^L\lambda_{jlk} \vect{R}_{jlk} \big\|_F^2}{\big( \frac1{\rho^{\rm{tr}}} + \big\|  \sum_{l=1}^L\vect{R}_{jlk} \big\|_2\big) \big( \frac1{\rho^{\rm{ul}}} + \big\|  \sum_{l=1}^L \big(\vect{R}_{jlk} - \vect{\Phi}_{jlk}\big) \big\|_2\big)  }
\end{align}
by applying Lemma~\ref{lemma2} twice. Notice that the denominator is bounded due to Assumption~\ref{assumption_4} and independent of $\{\lambda_{ljk}\}$. Therefore, if \eqref{Condition2_Assumption4_new} holds, it follows from \eqref{eq:AppendixG_99} that \eqref{Condition2_Assumption4_11} also holds.

We now exploit~\eqref{Condition2_Assumption4_11} to prove that $\vect{C}_{jk}$ is invertible for sufficiently large $M$. To this end, observe that $\vect{C}_{jk}$ with entries given by \eqref{eq:C_jk} is a Gramian matrix obtained as the inner products of the vectors $\{{\bf u}_{jlk} = \vecoperator \big(\frac{1}{\sqrt M}{\vect{Z}}_{j}^{-1/2}\vect{R}_{jlk} \vect{Q}_{jk}^{-1/2}\big):  \forall l\ne j\}$. Therefore, as $M$ grows large the matrix $\vect{C}_{jk}$ is invertible if and only if the vectors $\{{\bf u}_{jlk}: \forall l\ne j\} $ are asymptotically linearly independent.
Notice that the condition \eqref{Condition2_Assumption4_11} in Corollary~\ref{assumption_4_1}  for $\lalt=j$ can be rewritten in compact form as
\begin{align}\notag
\liminf_M \inf_{\{{\boldsymbol\lambda}_{jk}: \lambda_{jjk}=1\}}&\left({\bf u}_{jjk} + \sum\nolimits_{l\ne j}\lambda_{jlk}{\bf u}_{jlk}\right)^{\!\Htran}\\&\left({\bf u}_{jjk} + \sum\nolimits_{l\ne j}\lambda_{jlk}{\bf u}_{jlk}\right) > 0\label{eq:AppendixG_911}
\end{align}
which implies that the vectors $\{{\bf u}_{jlk}: \forall l\}$ are asymptotically linearly independent.
Since any subset of a finite set with linearly independent vectors is also linearly independent, \eqref{eq:AppendixG_911} ensures that $\{{\bf u}_{jlk}: \forall l\ne j\}$ are also asymptotically linearly independent. This proves that, under Assumption~\ref{assumption_4}, the Gramian matrix ${\bf C}_{jk}$ is invertible as $M\to \infty$ and this completes the proof.}

\bibliographystyle{IEEEtran}
\bibliography{IEEEabrv,ref}

\begin{IEEEbiography}
[{\includegraphics[width=1.0in,height=1.25in,clip,keepaspectratio]{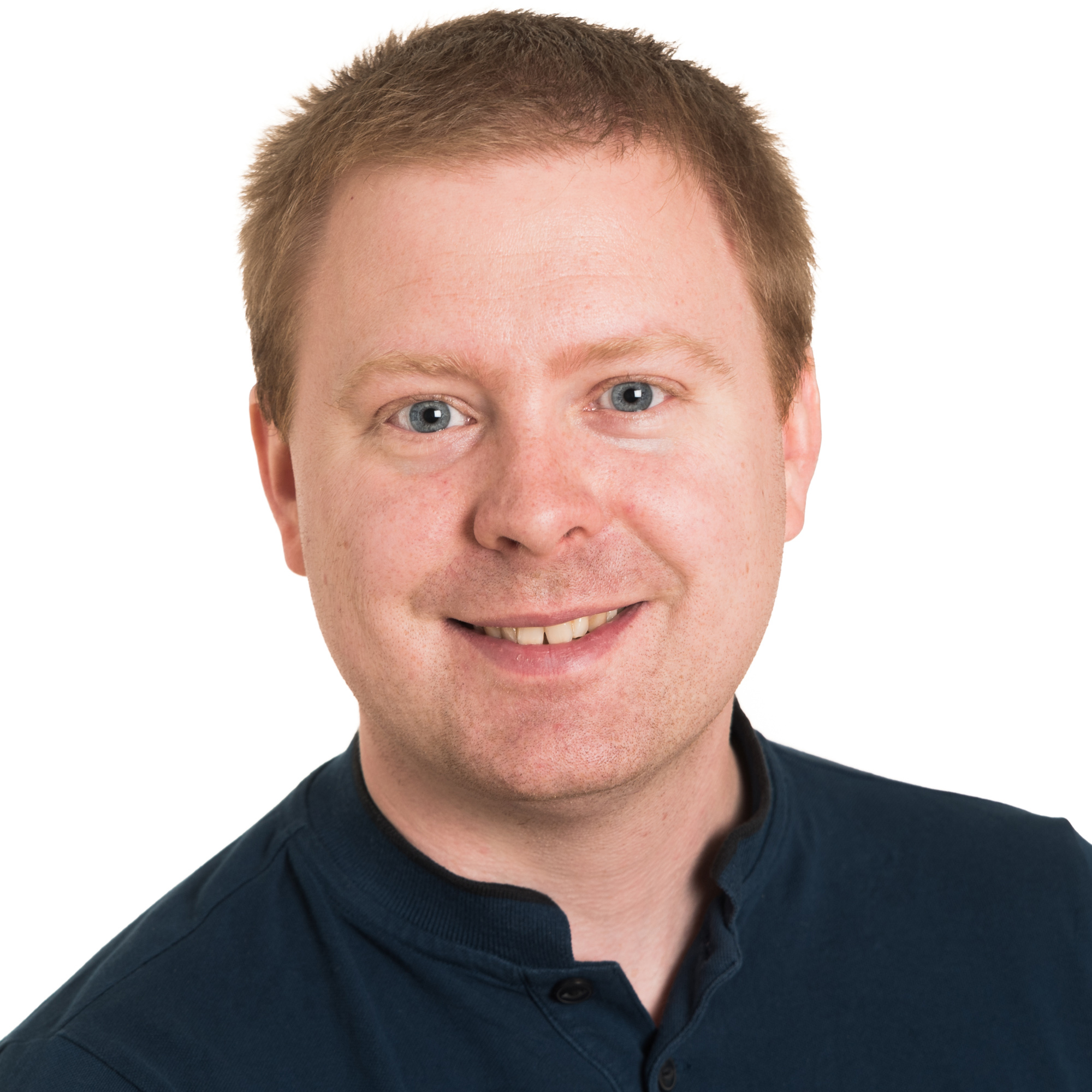}}]
{Emil Bj\"{o}rnson}(S'07, M'12) received the M.S. degree in Engineering Mathematics from Lund University, Sweden, in 2007. He received the Ph.D. degree in Telecommunications from KTH Royal Institute of Technology, Sweden, in 2011. From 2012 to mid 2014, he was a joint postdoc at the Alcatel-Lucent Chair on Flexible Radio, SUPELEC, France, and at KTH. He joined Linköping University, Sweden, in 2014 and is currently Senior Lecturer and Docent at the Division of Communication Systems.

He performs research on multi-antenna communications, Massive MIMO, radio resource allocation, energy-efficient communications, and network design. He is on the editorial board of the \textsc{IEEE Transactions on Communications} and the \textsc{IEEE Transactions on Green Communications and Networking}. He is the first author of the textbooks \emph{Massive MIMO Networks: Spectral, Energy, and Hardware Efficiency} (2017) and \emph{Optimal Resource Allocation in Coordinated Multi-Cell Systems} (2013). He is dedicated to reproducible research and has made a large amount of simulation code publicly available.

Dr. Björnson has performed MIMO research for more than ten years and has filed more than ten related patent applications. He received the 2016 Best PhD Award from EURASIP, the 2015 Ingvar Carlsson Award, and the 2014 Outstanding Young Researcher Award from IEEE ComSoc EMEA. He has co-authored papers that received best paper awards at WCSP 2017, IEEE ICC 2015, IEEE WCNC 2014, IEEE SAM 2014, IEEE CAMSAP 2011, and WCSP 2009.
\end{IEEEbiography}

\begin{IEEEbiography}[{\includegraphics[width=1in,height=1.25in,clip,keepaspectratio]{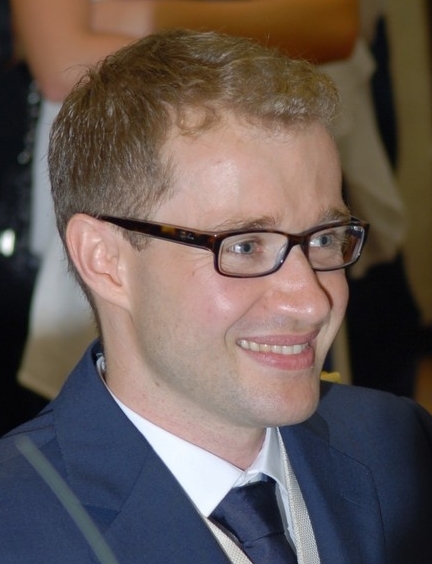}}]{Jakob Hoydis}(S'08--M'12) received the diploma degree (Dipl.-Ing.) in electrical engineering and information technology from RWTH Aachen University, Germany, and the Ph.D. degree from Sup\'{e}lec, Gif-sur-Yvette, France, in 2008 and 2012, respectively. He is a member of technical staff at Nokia Bell Labs, France, where he is investigating applications of deep learning for the physical layer. Previous to this position he was co-founder and CTO of the social network SPRAED and worked for Alcatel-Lucent Bell Labs in Stuttgart, Germany. His research interests are in the areas of machine learning, cloud computing, SDR, large random matrix theory, information theory, signal processing, and their applications to wireless communications. He is a co-author of the textbook \emph{Massive MIMO Networks: Spectral, Energy, and Hardware Efficiency} (2017). He is recipient of the 2012 Publication Prize of the Sup\'{e}lec Foundation, the 2013 VDE ITG F\"{o}rderpreis, and the 2015 Leonard G. Abraham Prize of the IEEE COMSOC. He received the IEEE WCNC 2014 best paper award and has been nominated as an Exemplary Reviewer 2012 for the IEEE Communication letters.
\end{IEEEbiography}

\begin{IEEEbiography}
[{\includegraphics[width=1.0in,height=1.25in,clip,keepaspectratio]{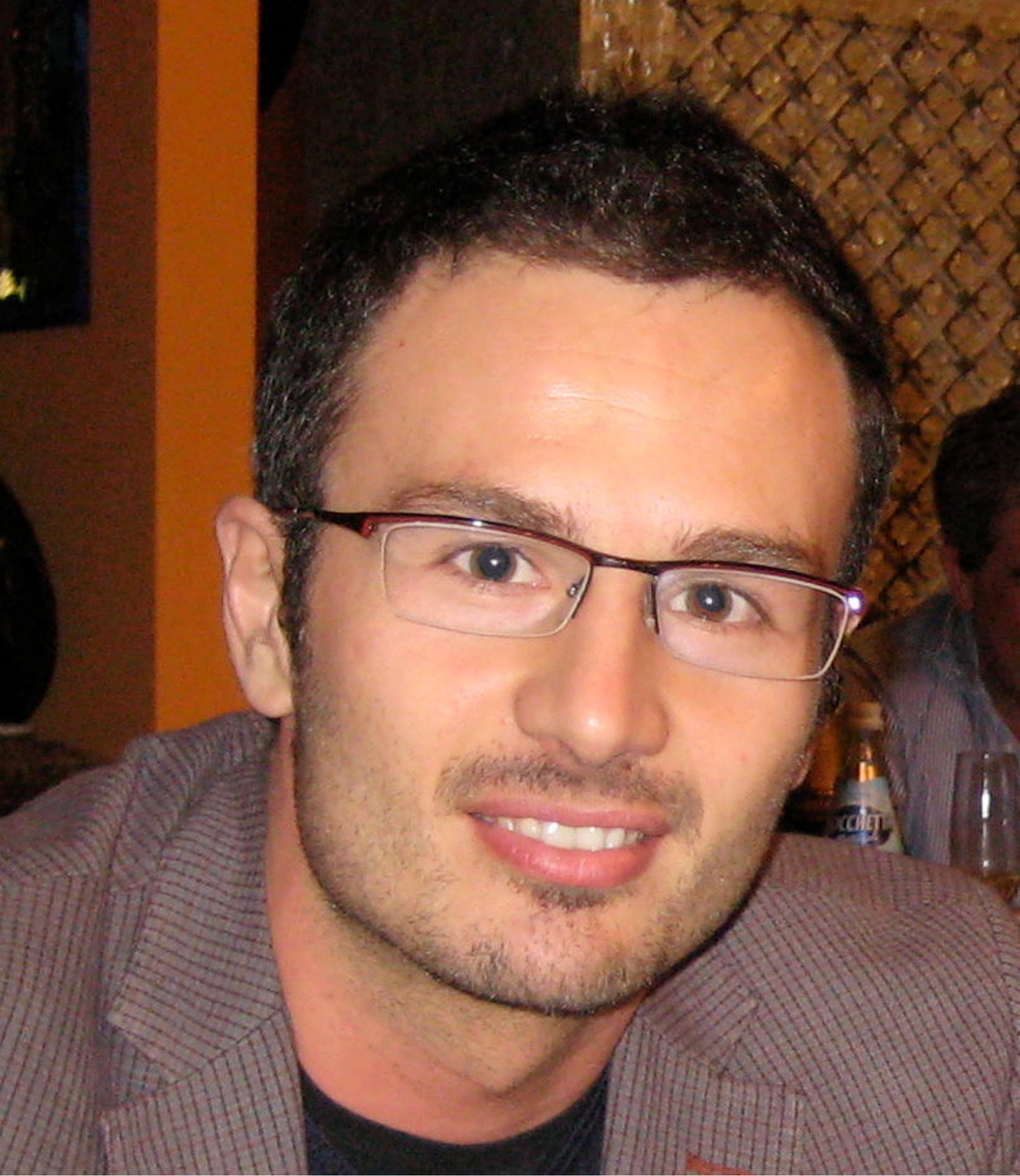}}]
{Luca Sanguinetti}(SM'15) received the Laurea Telecommunications Engineer degree (cum laude) and the Ph.D. degree in information engineering from the University of Pisa, Italy, in 2002 and 2005, respectively. Since 2005 he has been with the Dipartimento di Ingegneria dell'Informazione of the University of Pisa. In 2004, he was a visiting Ph.D. student at the German Aerospace Center (DLR), Oberpfaffenhofen, Germany. During the period June 2007 - June 2008, he was a postdoctoral associate in the Dept. Electrical Engineering at Princeton. During the period June 2010 - Sept. 2010, he was selected for a research assistantship at the Technische Universitat Munchen. From July 2013 to October 2017 he was with Large Systems and Networks Group (LANEAS), CentraleSup\'elec, Gif-sur-Yvette, France.

Dr. Sanguinetti is currently serving as an Associate Editor for the \textsc{IEEE Signal Processing Letters}. He served as an Associate Editor for \textsc{IEEE Transactions on Wireless communications}, and as Lead Guest Editor of \textsc{IEEE Journal on Selected Areas of Communications} Special Issue on ``Game Theory for Networks'' and as an Associate Editor for \textsc{IEEE Journal on Selected Areas of Communications} (series on Green Communications and Networking). Dr. Sanguinetti served as Exhibit Chair of the 2014 IEEE International Conference on Acoustics, Speech, and Signal Processing (ICASSP) and as the general co-chair of the 2016 Tyrrhenian Workshop on 5G\&Beyond. He is a co-author of the textbook \emph{Massive MIMO Networks: Spectral, Energy, and Hardware Efficiency} (2017).

His expertise and general interests span the areas of communications and signal processing, game theory and random matrix theory for wireless communications. He was the co-recipient of two best paper awards: \emph{IEEE Wireless Commun. and Networking Conference (WCNC) 2013} and \emph{IEEE Wireless Commun. and Networking Conference (WCNC) 2014}. He was also the recipient of the FP7 Marie Curie IEF 2013 ``Dense deployments for green cellular networks''. 
\end{IEEEbiography}

\end{document}